\documentclass[%
reprint,
superscriptaddress,
noeprint,
mismatch,amssymb,
prl,
longbibliography
]{revtex4-1}

\usepackage{hyperref}
\usepackage{graphicx}
\usepackage{dcolumn}
\usepackage{bm}
\usepackage{color}
\usepackage{braket}
\usepackage{here}
\usepackage{afterpage}

\usepackage{amsthm}
\usepackage{amsmath}	
\usepackage{upgreek}

\graphicspath{{./figs/}}

\begin{document}

\def\U#1{{%
\def\O{\mbox{O}}
\def\u{\mbox{u}}
\mathcode`\u=\mu
\mathcode`\O=\Omega
\mathrm{#1}}}
\def\ii{{\mathrm{i}}}
\def\jj{\,\mathrm{j}}                   
\def\ee{{\mathrm{e}}}
\def\dd{{\mathrm{d}}}
\def\cc{{\mathrm{c.c.}}}
\def\Re{\mathop{\mathrm{Re}}}
\def\Im{\mathop{\mathrm{Im}}}
\def\vct#1{\mathbf{#1}}
\def\fracpd#1#2{\frac{\partial#1}{\partial#2}}
\def\rank{\mathop{\mathrm{rank}}} 
\def\sub#1{_{\mbox{\scriptsize \rm #1}}}
\def\sur#1{^{\mbox{\scriptsize \rm #1}}}

\title{Topological Boundary Modes from Translational Deformations}

 \author{Yosuke Nakata}
\email{nakata@ee.es.osaka-u.ac.jp}
\affiliation{Research Center for Advanced Science and Technology (RCAST), The University of Tokyo, Meguro-ku, Tokyo 153-8904, Japan}
\affiliation{Graduate School of Engineering Science, Osaka University, Toyonaka,
Osaka 560-8531, Japan}
 \author{Yoshitaka Ito}
\affiliation{Research Center for Advanced Science and Technology (RCAST), The University of Tokyo, Meguro-ku, Tokyo 153-8904, Japan}
 \author{Yasunobu Nakamura}
 \affiliation{Research Center for Advanced Science and Technology (RCAST), The University of Tokyo, Meguro-ku, Tokyo 153-8904, Japan}
 \affiliation{Center for Emergent Matter Science (CEMS), RIKEN, Wako-shi, Saitama 351-0198, Japan}
\author{Ryuichi Shindou}
\affiliation{International Center for Quantum Materials, Peking University, Beijing 100871, China}
\affiliation{Collaborative Innovation Center of Quantum Matter, Beijing 100871, China}

\date{\today}

\begin{abstract}
Localized states universally appear when a periodic 
potential is perturbed by defects or terminated at its surface. 
In this Letter, we theoretically and experimentally demonstrate a 
mechanism that generates localized states through continuous 
translational deformations of periodic potentials. 
We provide a rigorous proof of 
the emergence of the localized states
under the deformations.
The mechanism is experimentally verified in microwave photonic crystals. 
We also demonstrate topological phase windings of reflected waves
for translated photonic crystals.
\end{abstract}



\maketitle
In the 1930s, Tamm predicted the localized state 
of an electron near the surface of a solid \cite{Tamm1932}. 
Years later, Shockley proposed another mechanism that produces
surface states, based on a band inversion of atomic orbitals \cite{Shockley1939}.
Impurities and lattice defects inside a crystal 
also produce localized states \cite{James1949, Saxon1949},
which play important roles in doped semiconductors. 
While such localized states were first investigated for electrons, they 
universally appear in various wave systems.
Zero-dimensional localized states have been observed in
electronic superlattices \cite{Ohno1990}, 
photonic and magnetophotonic crystals \cite{Yeh1978, Goto2008,Vinogradov2010},
plasmonic crystals \cite{Kitahara2003,Kitahara2004,Guo2008, Sasin2008, Dyer2013}, 
and phononic crystals \cite{Xiao2015}.

The recent discovery of topological insulators has shed fresh light 
on the understanding of surface states in various wave systems 
from a topological perspective. Under time-reversal symmetry, bulk electronic 
states in band insulators are generally characterized by the $Z_2$ 
topological invariant \cite{Kane2005,Fu2007}. The bulk-edge correspondence 
relates the bulk $Z_2$ topological invariant to surface characteristics 
and ensures an existence of gapless boundary states with the Kramers degeneracy 
protected by time-reversal symmetry \cite{Konig2007,Chen2009}. Later, it was shown 
that other discrete symmetries and their combinations generate various topological 
numbers for 
bulk electronic states and associated in-gap gapless boundary states \cite{Schnyder2008}. 
A pioneering example is the $Z$ topological invariant with a sublattice symmetry in 
the Su-Schrieffer-Heeger model \cite{Su1979, Su1980}. 
The nonzero topological integer in the Su-Schrieffer-Heeger model ensures zero-energy end states with 
sublattice-symmetry protection.  For continuous one-dimensional crystals with 
inversion symmetry, Xiao \textit{et~al.} established a relation between surface 
observables and  bulk properties and rigorously determined the existence or 
nonexistence of localized states \cite{Xiao2014}.  So far, research on one-dimensional 
systems has focused on unit cells with either sublattice or inversion symmetry to 
define the topological integers, but these discrete symmetries may not be essential,
as suggested by Shockley \cite{Shockley1939}. In fact, the in-gap localized states as 
boundary states could survive under a gradual structural deformation that breaks the 
symmetries within the unit cell.  This consideration indicates an alternative topological 
mechanism that generates localized states without using any symmetry protection. 

In this letter, we devise a scheme that produces zero-dimensional localized states 
in a defect created by a translational deformation of a periodic potential.
A rigorous proof of emergence of the localized states is provided without relying 
on any symmetry protection. The scheme is experimentally demonstrated in 
microwave photonic crystals.

Consider a one-particle eigenmode in one-dimensional 
continuous media with a periodic potential of the period $a$. 
From the Bloch theorem~\cite{ashcroft1976solid},
the eigenmode $\ket{\psi_n(k)}= \exp(i k \hat{x}) \ket{u_n(k)}$ 
is characterized by the crystal momentum $k$ in the first Brillouin zone 
$[-\pi/a,\pi/a]$  and the energy band index $n$, where
$\braket{x|u_n(k)}$ is periodic in $x$.
For simplicity, we assume that 
the eigenenergy $E_n(k)$ of 
$\ket{\psi_n(k)}$ satisfies
$E_1(k) < E_2(k) < \cdots$ in the entire Brillouin zone.
From here, we focus on the $n$th band.
The first Brillouin zone is discretized as 
$k_m = m \pi/(Ma)$ with $m=-M+1, -M+2,..., M$ ($2M$ points). 
In terms of $\ket{u_m} = \ket{u_n(k_m)}$, a Wilson loop is 
given by
\begin{multline}
  W = \braket{u_M|u_{M-1}}  \braket{u_{M-1}|u_{M-2}}
 \cdots \braket{u_{1}|u_{0}} \times \\
\braket{u_{0}|u_{-1}}\cdots\braket{u_{-M+2}|u_{-M+1}}\braket{u_{-M+1}|e^{i G\hat{x}}|u_{M}},  \label{eq:1}
\end{multline}
where $G=2\pi/a$ and the inner product is defined in the unit cell \cite{Vanderbilt2018}.
It is normalized to be unity 
as $\lim_{M\rightarrow \infty} 
W = \exp(i \theta\sub{Zak})$, where 
$\theta\sub{Zak}$ is simply 
the Zak phase \cite{Zak1989}. The Zak phase specifies a spatial 
displacement of the localized Wannier orbits that 
are composed only of the eigenmodes in the $n$th energy band~\cite{Vanderbilt2018}.  
In electronic systems, the Zak phase corresponds to surface charge,
which can take a fractional value 
\cite{Vanderbilt1993,Gangadharaiah2012,Park2016,Thakurathi2018}.

\begin{figure}[tbph]
 \centering
  \includegraphics{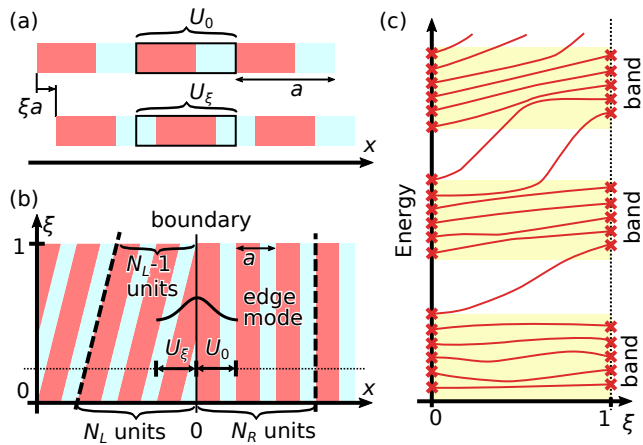}
  \caption{(a)~One-dimensional continuous medium with a periodic potential (upper) and the 
identical system with spatial translation by $\xi a$ (lower).
(b)~Spatial boundary between the two identical periodic systems with different translations. 
The Born-von-Karman (BvK) boundary condition is imposed, where the two dashed lines  
are identified with each other.
(c)~Eigenenergies of the entire system with the BvK 
boundary condition are schematically plotted as a function of the 
translational parameter $\xi$. 
\label{fig:translational_pumping}}
 \end{figure}
Now, let us translate continuously the one-dimensional periodic potential by $\xi a$ 
($0\leq \xi \leq 1$) relative to a fixed frame of the unit cell.  
The spatial translation changes the potential configuration from $U_0$ to $U_\xi$ inside the fixed cell
[see Fig.~\ref{fig:translational_pumping}(a)]. 
When $\xi$ changes from 0 to 1, 
the localized Wannier orbit is continuously translated by the periodic 
length $a$.
Thus, it works like a classical screw pump \cite{Ozawa2019}.
Being identical to the displacement of the Wannier 
orbit with the unit-cell length (apart from a factor $2\pi/a$), 
the Zak phase also continuously increases 
by $2\pi$ under the translation:
$\int^{1}_{0} \dd\xi\, \partial_{\xi} \theta\sub{Zak} = 2\pi$.
The phase winding counts 
the Chern integer, 
which represents the topological characteristics of a fiber bundle on 
the $(k,\xi)$ plane \cite{Vanderbilt2018, asboth2016short}. 
In this Letter,  $\xi$  is regarded as a variable independent of other
variables. Nonetheless, one could consider a continuous change of $\xi$ as
a function of time $t$. In particular, an adiabatic change of $\xi=\xi(t)$ from 0 to 1
in $t$ suppresses interband transitions and is referred to as Thouless pumping \cite{Thouless1983}.

The $2\pi$ phase winding in the Zak phase under the translation 
leads to a series of nontrivial localized states in a spatial boundary  
between two identical one-dimensional periodic systems with different translations 
$\xi$. To see this, let us consider a periodic arrangement of 
unit cells with $U_\xi$ in a region of $x< 0$ 
and another periodic arrangement of unit cells with $U_0$ in the other 
region of $x\geq 0$. The translation parameter $\xi$ and spatial coordinate $x$ subtend 
an extended two-dimensional space, as shown in Fig.~\ref{fig:translational_pumping}(b).
When $\xi$ changes from 0 to 1, the Zak phase in the former bulk region ($x < 0$) winds up 
the $2\pi$ phase. Meanwhile, the Zak phase in the latter bulk region ($x \geq 0$) remains unchanged.
Accordingly, the bulk-edge correspondence 
\cite{Thouless1983, asboth2016short, Vanderbilt2018} suggests 
the existence of zero-dimensional edge states at the boundary region ($x=0$), 
whose eigenenergies have ``chiral'' dispersions within a bulk band gap as a 
function of the translational parameter 
$\xi$ [Fig.~\ref{fig:translational_pumping}(c)]. Moreover, as the Zak phase for 
any bulk band in the region of $x < 0$ acquires the same $2\pi$ phase winding during the 
translation, the number of the chiral dispersions 
between the $n$th and $(n+1)$th bulk bands are 
expected to be $n$ [Fig.~\ref{fig:translational_pumping}(c)].  

To prove this bulk-edge correspondence in the translational deformation
rigorously, let us impose the following Born-von-Karman (BvK) boundary condition 
on a finite system [Fig.~\ref{fig:translational_pumping}(b)]. Suppose that at $\xi=0$, the entire 
one-dimensional system is comprised of $N_L$ unit cells in the region of $x< 0$ and 
$N_R$ unit cells in the region of $x\geq 0$. 
For general $\xi$, we identify $x=-N_La + \xi a$ with $x=N_R a$, such that 
the lattice periodicity is preserved at $x=N_Ra\equiv -N_La+\xi a$ and it is broken only at $x=0$. 
For $\xi=0$ and $\xi=1$, the periodicity is completely preserved in the entire system, so that 
the eigenmodes at $\xi=0$ and $\xi=1$ are all spatially extended bulk band states.
Under the BvK boundary condition, which discretizes the Brillouin zone,
numbers of the bulk modes in each band at $\xi=0$ and at $\xi=1$ 
are given by $N_L+N_R$ and $N_L+N_R-1$, respectively.
Namely, the number of the extended bulk states 
decreases by one in each band when $\xi$ continuously changes from $0$ to $1$. 
As the energy has a lower bound and 
there is no upper bound on the bulk band index $n$
in continuous media,
more than one eigenmode in each bulk band at $\xi=0$ 
must move into bulk bands with a higher energy at $\xi=1$ 
during the translation of $\xi$. 
For example, when one eigenmode in the lowest bulk band at $\xi=0$ 
goes to the second lowest bulk band at $\xi=1$, two eigenmodes in the second 
lowest band at $\xi=0$ must go to the third lowest one at $\xi=1$ 
[Fig.~\ref{fig:translational_pumping}(c)]. 
This argument inductively dictates that during the translation 
of $\xi$, $n$ modes always raise their energies out of the $n$th bulk energy band 
and go across the band gap between the $n$th and $(n+1)$th bands. 
An in-gap mode generally has a 
complex-valued wave number \cite{Cottey1971}. Accordingly, the $n$ in-gap modes must 
be spatially localized at $x=0$, where the lattice periodicity is broken; 
therefore, they are simply  
defect modes localized at the boundary.  
Importantly, the argument so far 
does not require
any symmetry protection
for the presence of the in-gap localized states.

\begin{figure}[!t]
 \centering
  \includegraphics{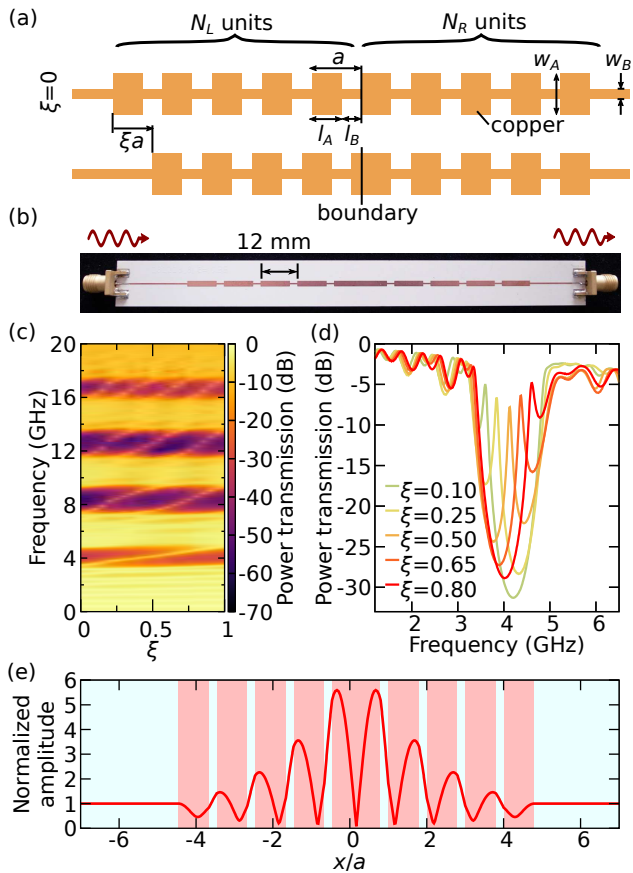}
  \caption{(a)~Schematic top view of binary
microstrips at $\xi=0$ (upper) and at $\xi\ne 0$ (lower). 
(b)~Photograph of a sample with $\xi = 0.35$. 
The structural parameters are $a=12\,\U{mm}$, $l_A/a=0.8$, $l_B/a=0.2$, 
$w_A=1.8\,\U{mm}$, $w_B = 0.45\,\U{mm}$, and $N_L=N_R=5$.
The microstrip is made of a $35\,\mu\U{m}$-thick copper film 
on a polyphenylene-ether substrate (\textsc{Risho} CS-3396; thickness $0.56\,\U{mm}$,
$\epsilon=11.3$, $\tan \delta = 0.003$ at $1\,\U{GHz}$),
and SMA connectors (\textsc{GigaLane} PSF-S01-001) are attached to the substrate.
The back of the substrate contains a ground plane made of a copper film with the same
thickness as the microstrip.
(c)~Power transmission spectra through the 21 samples
from $\xi=0$ to $\xi=1$ with the step size of $\Delta \xi=0.05$.
The input power is set to $0\,\U{dBm}$.
(d)~Power transmission spectra inside the first band gap for several values of $\xi$.
(e)~Calculated distribution of the absolute value of the complex electric-field 
amplitude at $4.22\,\U{GHz}$ with $\xi=0.538$ inside the first band gap. 
The amplitude is normalized to that of the incident field. 
Regions of different colors represent different width strips.
\label{fig:exp_trans}}
 \end{figure}

Now, we experimentally confirm the theoretical concept by using microstrip 
photonic crystals. A microstrip is a transmission line composed of a metallic strip 
separated from a conducting ground plane by a dielectric substrate.
Microwaves propagate between the topside metallic strip and the backside ground plane,
and the impedance and refractive index of a microstrip are determined 
by the geometrical parameters. 

The first photonic system studied has a binary unit cell,
in which the two strips with different widths behave as two different media.  
As shown schematically in Fig.~\ref{fig:exp_trans}(a), 
we continuously introduce a defect around the boundary 
by displacing the left half by $\xi a$ while leaving the right half unchanged.
A photograph of one of the fabricated samples ($\xi=0.35$) is provided
in Fig.~\ref{fig:exp_trans}(b).
Using a vector network analyzer (\textsc{Keysight} 5232A),
we measured the power transmission 
through the samples with $\xi$ from $0$ to $1$ 
with a step size of $\Delta \xi=0.05$. 
The transmission spectra obtained for these different $\xi$ are summarized 
in Fig.~\ref{fig:exp_trans}(c). Under $20\,\U{GHz}$, we clearly see 
five transmission bands, and four band gaps between them. 
The $n$th band gap has $n$ boundary modes 
running between the two neighboring transmission bands, as expected from 
the theory. The qualitative behavior of the transmission spectra can be well captured 
by a transfer-matrix model calculation~\cite{supplemental}. 
Figure~\ref{fig:exp_trans}(d) shows some of 
the experimentally obtained power transmission spectra inside the first 
band gap. The transmission peak decreases and 
the line width becomes narrower around $\xi=0.50$. This is  
because coupling between the incident wave 
and the boundary mode is reduced at the center of the band gap. In fact, 
the transfer-matrix model calculation confirms that 
the localized mode becomes the narrowest at the center of 
the first band gap~\cite{supplemental}, as plotted in Fig.~\ref{fig:exp_trans}(e).

\begin{figure}[!t]
 \centering
  \includegraphics{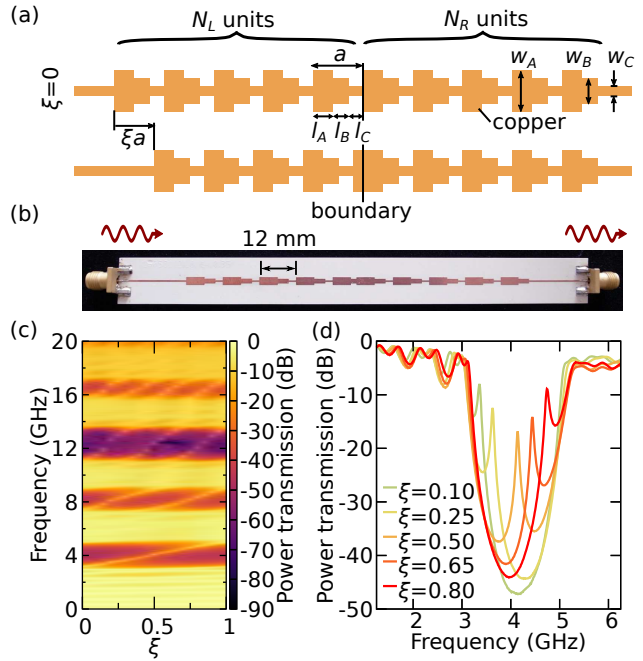}
  \caption{(a)~Schematic top view of ternary
microstrips at $\xi=0$ (upper) and at $\xi\ne 0$ (lower). 
(b)~Photograph of a sample with $\xi = 0.35$. 
The structural parameters are $l_A/a=0.5$, $l_B/a=0.3$,
$l_C/a=0.2$, $w_A=2.5\,\U{mm}$, $w_B=1.5\,\U{mm}$, and
$w_C=0.45\,\U{mm}$. The other parameters are
the same as those in Fig.~\ref{fig:exp_trans}.
(c)~Power transmission spectra through the 21 samples
from $\xi=0$ to $\xi=1$ with a step size of $\Delta \xi=0.05$.
(d)~Power transmission spectra inside the first band gap for several values of $\xi$.
 \label{fig:exp_trans_without_inversion}
}
 \end{figure}

The second photonic system studied has three components in the unit cell. 
The design and photograph of the ternary microstrips are shown in 
Figs.~\ref{fig:exp_trans_without_inversion}(a) and \ref{fig:exp_trans_without_inversion}(b), respectively.
With three different regions, 
the unit cell has no spatial inversion symmetry at any $\xi$. 
Figures~\ref{fig:exp_trans_without_inversion}(c) and \ref{fig:exp_trans_without_inversion}(d) illustrate 
the experimental transmission spectra for 
21 samples with different $\xi$~\cite{supplemental}.
These transmission spectra confirm that the $n$ localized modes run across 
the $n$th transmission gap during the translation of $\xi$ from $0$ to $1$. The experimental 
results clearly demonstrate that inversion symmetry is not essential for the generation 
of the series of localized states through translational deformation.

\begin{figure}[tbp]
 \centering
  \includegraphics{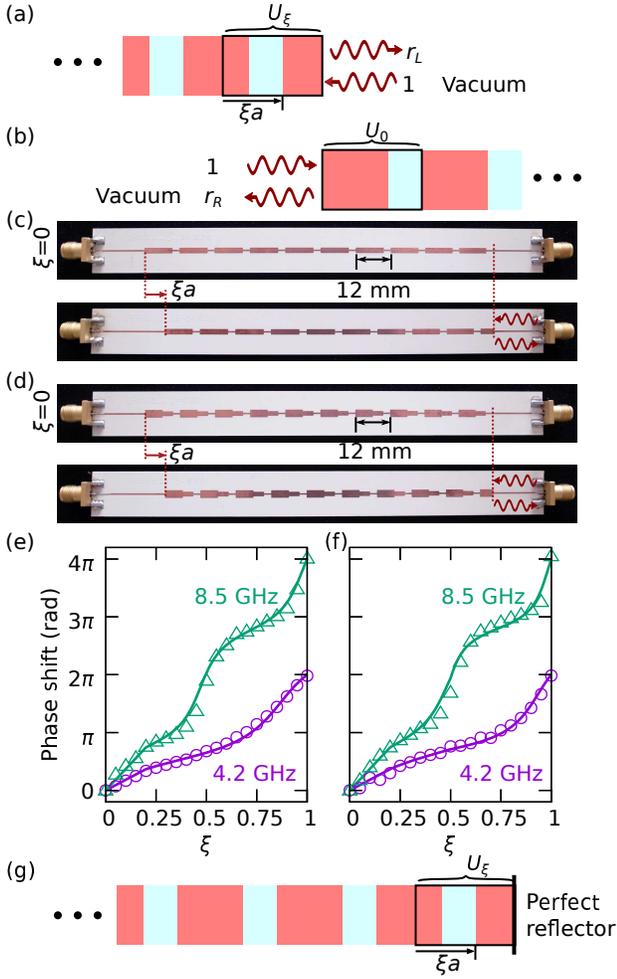}
  \caption{
Definitions of (a)~left and (b)~right 
complex reflection amplitudes $r_L(\xi,\omega)$ 
and $r_R(\omega)$, respectively.  
Photograph of (c) binary and (d) ternary samples ($\xi=0$ and $0.60$) 
to measure $r_L(\xi,\omega)$.
The parameters are the same as those in Figs.~\ref{fig:exp_trans}
and~\ref{fig:exp_trans_without_inversion}.
Topological winding of
$\arg[r_L(\xi, \omega_0)] - \arg[r_L(0, \omega_0)]$
as a function of $\xi$ 
for (e) binary and (f) ternary samples.
Here, $\omega_0$ is set to $2\pi\times 4.20\,\U{GHz}$ and
$2\pi\times 8.50\,\U{GHz}$
inside the first band gap (circles) and second band gap (triangles), respectively.
A microwave is injected from the right connector; meanwhile,
the left connector is connected to 
another port of the network analyzer through a cable.
The theoretical curves for the semi-infinite systems are plotted with the experimentally obtained points.
Note that we use the convention of $\arg\, e^{j \theta}=\theta$ for $j=-i$.
(g)~One-dimensional photonic crystal terminated by a perfect reflector. 
 \label{fig:reflection_phase_winding}
}
\end{figure}

Next, we establish 
the physical origin of the localized states in terms of  
phase winding of the complex reflection amplitude. 
To this end, we divide the deformed crystal into two halves. 
Namely, the left region with $U_{\xi}$ is now terminated 
at its right end by a vacuum region, while the right region with 
$U_0$ is terminated by the same vacuum region at its left end 
[Figs.~\ref{fig:reflection_phase_winding}(a) and \ref{fig:reflection_phase_winding}(b)]. 
Photonic properties of each semi-infinite region are characterized 
by the complex reflection amplitude $r$ or 
a relative surface impedance $Z^{(s)} \equiv (1+r)/(1-r)$ 
at the respective termination. 
The topological characteristics of localized states are encoded  
between the complex reflection amplitudes at both terminations  
$r_L(\xi, \omega)$ and $r_R(\omega)$ 
with an angular frequency $\omega$. 
Specifically, a condition for eigenmodes localized at the 
original defect is nothing but the resonance condition across the two 
terminations: $Z_L^{(s)}(\xi,\omega) + Z_R^{(s)}(\omega)= 0$.  
The resonance condition can also be written as  
$r_L(\xi,\omega) r_R(\omega)=1$. 
When $\omega$ remains inside the band gap between the $n$th and $(n+1)$th 
bulk bands, both the semi-infinite regions behave as perfect reflectors: 
$|r_R(\omega)|=|r_L(\xi, \omega)|=1$ under assumption of no dissipation. 
Thus, the condition shows that the phase of $r_L(\xi,\omega)$ 
must wind up by $2\pi n$ during the translation from $\xi=0$ to 
$\xi=1$, because the $n$ boundary modes move 
across the angular frequency $\omega$ in the gap. The direction of the winding 
is determined by Foster's theorem \cite{Collin1996,supplemental}.

The $2\pi n$ phase winding of the reflection is considered 
as the physical origin of the localized states.
To confirm this phase winding experimentally, we fabricated samples 
composed only of the left-half parts with different $\xi$, 
as in Figs.~\ref{fig:reflection_phase_winding}(c) and \ref{fig:reflection_phase_winding}(d).
Figures~\ref{fig:reflection_phase_winding}(e) and \ref{fig:reflection_phase_winding}(f)
show measured phases of the reflected waves of samples with different 
$\xi$ (relative to the measured phase at $\xi=0$). 
The experimental data points agree well with the theoretical curves obtained 
from the transfer-matrix model calculations for the semi-infinite systems~\cite{supplemental}.
The results clearly demonstrate the presence of phase winding of the 
reflection amplitude, regardless of the unit-cell symmetry.

The phase winding of the reflection provides a unified perspective 
on 
both Tamm and Shockley states, which are often separately 
attributed to a perturbed surface potential and band inversion,
respectively \cite{Tamm1932,Shockley1939,Vinogradov2010}.
To this end, we consider that the left region with $U_{\xi}$ is terminated 
by a perfect reflector at the right end
as shown in Fig.~\ref{fig:reflection_phase_winding}(g).
Given $|r_R(\omega)|=1$ for those $\omega$ in the $n$th 
transmission gap of the left part, the $2\pi n$ phase winding of 
$r_L(\xi,\omega)$ during the translation of $\xi$ from 
0 to 1 always guarantees 
the emergence of $n$ localized eigenmodes at the 
termination, irrespective of the details of the reflector on the right side. 
This holds true for any reflector with $\xi$-independent perturbations, 
provided the perturbations maintain the perfect-reflection condition.
Such perturbations include a delta-function-like surface
perturbation, the existence of which distinguishes Tamm states from Shockley states,
as discussed in 
Ref.~\cite{Shockley1939}.
In this sense, our proposed mechanism provides a comprehensive viewpoint for 
both Tamm and Shockley states.

In summary, we demonstrated a scenario that produces localized states  
through translational deformations analogous to classical screw pumping. 
The mechanism is not restricted 
to a specific physical system; rather, it is universal for any waves.
Localized states in a system, even in the absence of sublattice or inversion symmetry,
are now interpreted as topological boundary  modes. The termination at the spatial 
boundary is understood as an engineered degree of freedom and can be used 
for tuning the spatial localization of the boundary mode.

The authors thank K.~Usami, A.~Noguchi, and M.~W.~Takeda for their fruitful discussions,
and J.~Koenig for his careful reading of the manuscript.
This work was supported by JSPS KAKENHI (Grant No.~17K17777) 
and by the JST ERATO project (Grant No.~JPMJER1601). R.~S. was supported 
by National Basic Research Programs of China (973 program Grants No.~2014CB920901 and No.~2015CB921104)
and National Natural Science Foundation of China (Grants No.~2017A040215).


\begin{thebibliography}{37}%
\makeatletter
\providecommand \@ifxundefined [1]{%
 \@ifx{#1\undefined}
}%
\providecommand \@ifnum [1]{%
 \ifnum #1\expandafter \@firstoftwo
 \else \expandafter \@secondoftwo
 \fi
}%
\providecommand \@ifx [1]{%
 \ifx #1\expandafter \@firstoftwo
 \else \expandafter \@secondoftwo
 \fi
}%
\providecommand \natexlab [1]{#1}%
\providecommand \enquote  [1]{``#1''}%
\providecommand \bibnamefont  [1]{#1}%
\providecommand \bibfnamefont [1]{#1}%
\providecommand \citenamefont [1]{#1}%
\providecommand \href@noop [0]{\@secondoftwo}%
\providecommand \href [0]{\begingroup \@sanitize@url \@href}%
\providecommand \@href[1]{\@@startlink{#1}\@@href}%
\providecommand \@@href[1]{\endgroup#1\@@endlink}%
\providecommand \@sanitize@url [0]{\catcode `\\12\catcode `\$12\catcode
  `\&12\catcode `\#12\catcode `\^12\catcode `\_12\catcode `\%12\relax}%
\providecommand \@@startlink[1]{}%
\providecommand \@@endlink[0]{}%
\providecommand \url  [0]{\begingroup\@sanitize@url \@url }%
\providecommand \@url [1]{\endgroup\@href {#1}{\urlprefix }}%
\providecommand \urlprefix  [0]{URL }%
\providecommand \Eprint [0]{\href }%
\providecommand \doibase [0]{https://doi.org/}%
\providecommand \selectlanguage [0]{\@gobble}%
\providecommand \bibinfo  [0]{\@secondoftwo}%
\providecommand \bibfield  [0]{\@secondoftwo}%
\providecommand \translation [1]{[#1]}%
\providecommand \BibitemOpen [0]{}%
\providecommand \bibitemStop [0]{}%
\providecommand \bibitemNoStop [0]{.\EOS\space}%
\providecommand \EOS [0]{\spacefactor3000\relax}%
\providecommand \BibitemShut  [1]{\csname bibitem#1\endcsname}%
\let\auto@bib@innerbib\@empty
\bibitem [{\citenamefont {Tamm}(1932)}]{Tamm1932}%
  \BibitemOpen
  \bibfield  {author} {\bibinfo {author} {\bibfnamefont {I.}~\bibnamefont
  {Tamm}},\ }\bibfield  {title} {\bibinfo {title} {{{\"{U}}ber eine
  m{\"{o}}gliche art der elektronenbindung an kristalloberfl{\"{a}}chen}},\
  }\href@noop {} {\bibfield  {journal} {\bibinfo  {journal} {Phys. Z.
  Sowjetunion}\ }\textbf {\bibinfo {volume} {1}},\ \bibinfo {pages} {733}
  (\bibinfo {year} {1932})}\BibitemShut {NoStop}%
\bibitem [{\citenamefont {Shockley}(1939)}]{Shockley1939}%
  \BibitemOpen
  \bibfield  {author} {\bibinfo {author} {\bibfnamefont {W.}~\bibnamefont
  {Shockley}},\ }\bibfield  {title} {\bibinfo {title} {{On the surface states
  associated with a periodic potential}},\ }\href
  {https://doi.org/10.1103/PhysRev.56.317} {\bibfield  {journal} {\bibinfo
  {journal} {Phys. Rev.}\ }\textbf {\bibinfo {volume} {56}},\ \bibinfo {pages}
  {317} (\bibinfo {year} {1939})}\BibitemShut {NoStop}%
\bibitem [{\citenamefont {James}(1949)}]{James1949}%
  \BibitemOpen
  \bibfield  {author} {\bibinfo {author} {\bibfnamefont {H.~M.}\ \bibnamefont
  {James}},\ }\bibfield  {title} {\bibinfo {title} {{Electronic states in
  perturbed periodic systems}},\ }\href@noop {} {\bibfield  {journal} {\bibinfo
   {journal} {Phys. Rev.}\ }\textbf {\bibinfo {volume} {76}},\ \bibinfo {pages}
  {1611} (\bibinfo {year} {1949})}\BibitemShut {NoStop}%
\bibitem [{\citenamefont {Saxon}\ and\ \citenamefont
  {Hunter}(1949)}]{Saxon1949}%
  \BibitemOpen
  \bibfield  {author} {\bibinfo {author} {\bibfnamefont {D.~S.}\ \bibnamefont
  {Saxon}}\ and\ \bibinfo {author} {\bibfnamefont {R.~A.}\ \bibnamefont
  {Hunter}},\ }\bibfield  {title} {\bibinfo {title} {{Some electronic
  properties of a one-dimensional crystal model}},\ }\href@noop {} {\bibfield
  {journal} {\bibinfo  {journal} {Philips Res. Rep.}\ }\textbf {\bibinfo
  {volume} {4}},\ \bibinfo {pages} {81} (\bibinfo {year} {1949})}\BibitemShut
  {NoStop}%
\bibitem [{\citenamefont {Ohno}\ \emph {et~al.}(1990)\citenamefont {Ohno},
  \citenamefont {Mendez}, \citenamefont {Brum}, \citenamefont {Hong},
  \citenamefont {Agull{\'{o}}-Rueda}, \citenamefont {Chang},\ and\
  \citenamefont {Esaki}}]{Ohno1990}%
  \BibitemOpen
  \bibfield  {author} {\bibinfo {author} {\bibfnamefont {H.}~\bibnamefont
  {Ohno}}, \bibinfo {author} {\bibfnamefont {E.~E.}\ \bibnamefont {Mendez}},
  \bibinfo {author} {\bibfnamefont {J.~A.}\ \bibnamefont {Brum}}, \bibinfo
  {author} {\bibfnamefont {J.~M.}\ \bibnamefont {Hong}}, \bibinfo {author}
  {\bibfnamefont {F.}~\bibnamefont {Agull{\'{o}}-Rueda}}, \bibinfo {author}
  {\bibfnamefont {L.~L.}\ \bibnamefont {Chang}},\ and\ \bibinfo {author}
  {\bibfnamefont {L.}~\bibnamefont {Esaki}},\ }\bibfield  {title} {\bibinfo
  {title} {{Observation of ``Tamm states'' in Superlattices}},\ }\href
  {https://doi.org/10.1103/PhysRevLett.64.2555} {\bibfield  {journal} {\bibinfo
   {journal} {Phys. Rev. Lett.}\ }\textbf {\bibinfo {volume} {64}},\ \bibinfo
  {pages} {2555} (\bibinfo {year} {1990})}\BibitemShut {NoStop}%
\bibitem [{\citenamefont {Yeh}\ \emph {et~al.}(1978)\citenamefont {Yeh},
  \citenamefont {Yariv},\ and\ \citenamefont {Cho}}]{Yeh1978}%
  \BibitemOpen
  \bibfield  {author} {\bibinfo {author} {\bibfnamefont {P.}~\bibnamefont
  {Yeh}}, \bibinfo {author} {\bibfnamefont {A.}~\bibnamefont {Yariv}},\ and\
  \bibinfo {author} {\bibfnamefont {A.~Y.}\ \bibnamefont {Cho}},\ }\bibfield
  {title} {\bibinfo {title} {{Optical surface waves in periodic layered
  media}},\ }\href {https://doi.org/10.1063/1.89953} {\bibfield  {journal}
  {\bibinfo  {journal} {Appl. Phys. Lett.}\ }\textbf {\bibinfo {volume} {32}},\
  \bibinfo {pages} {104} (\bibinfo {year} {1978})}\BibitemShut {NoStop}%
\bibitem [{\citenamefont {Goto}\ \emph {et~al.}(2008)\citenamefont {Goto},
  \citenamefont {Dorofeenko}, \citenamefont {Merzlikin}, \citenamefont
  {Baryshev}, \citenamefont {Vinogradov}, \citenamefont {Inoue}, \citenamefont
  {Lisyansky},\ and\ \citenamefont {Granovsky}}]{Goto2008}%
  \BibitemOpen
  \bibfield  {author} {\bibinfo {author} {\bibfnamefont {T.}~\bibnamefont
  {Goto}}, \bibinfo {author} {\bibfnamefont {A.~V.}\ \bibnamefont
  {Dorofeenko}}, \bibinfo {author} {\bibfnamefont {A.~M.}\ \bibnamefont
  {Merzlikin}}, \bibinfo {author} {\bibfnamefont {A.~V.}\ \bibnamefont
  {Baryshev}}, \bibinfo {author} {\bibfnamefont {A.~P.}\ \bibnamefont
  {Vinogradov}}, \bibinfo {author} {\bibfnamefont {M.}~\bibnamefont {Inoue}},
  \bibinfo {author} {\bibfnamefont {A.~A.}\ \bibnamefont {Lisyansky}},\ and\
  \bibinfo {author} {\bibfnamefont {A.~B.}\ \bibnamefont {Granovsky}},\
  }\bibfield  {title} {\bibinfo {title} {{Optical Tamm States in
  One-Dimensional Magnetophotonic Structures}},\ }\href
  {https://doi.org/10.1103/PhysRevLett.101.113902} {\bibfield  {journal}
  {\bibinfo  {journal} {Phys. Rev. Lett.}\ }\textbf {\bibinfo {volume} {101}},\
  \bibinfo {pages} {113902} (\bibinfo {year} {2008})}\BibitemShut {NoStop}%
\bibitem [{\citenamefont {Vinogradov}\ \emph {et~al.}(2010)\citenamefont
  {Vinogradov}, \citenamefont {Dorofeenko}, \citenamefont {Merzlikin},\ and\
  \citenamefont {Lisyansky}}]{Vinogradov2010}%
  \BibitemOpen
  \bibfield  {author} {\bibinfo {author} {\bibfnamefont {A.~P.}\ \bibnamefont
  {Vinogradov}}, \bibinfo {author} {\bibfnamefont {A.~V.}\ \bibnamefont
  {Dorofeenko}}, \bibinfo {author} {\bibfnamefont {A.~M.}\ \bibnamefont
  {Merzlikin}},\ and\ \bibinfo {author} {\bibfnamefont {A.~A.}\ \bibnamefont
  {Lisyansky}},\ }\bibfield  {title} {\bibinfo {title} {{Surface states in
  photonic crystals}},\ }\href {https://doi.org/10.3367/UFNe.0180.201003b.0249}
  {\bibfield  {journal} {\bibinfo  {journal} {Phys. Usp.}\ }\textbf {\bibinfo
  {volume} {53}},\ \bibinfo {pages} {243} (\bibinfo {year} {2010})}\BibitemShut
  {NoStop}%
\bibitem [{\citenamefont {Kitahara}\ \emph {et~al.}(2003)\citenamefont
  {Kitahara}, \citenamefont {Kawaguchi}, \citenamefont {Miyashita},\ and\
  \citenamefont {{Wada Takeda}}}]{Kitahara2003}%
  \BibitemOpen
  \bibfield  {author} {\bibinfo {author} {\bibfnamefont {H.}~\bibnamefont
  {Kitahara}}, \bibinfo {author} {\bibfnamefont {T.}~\bibnamefont {Kawaguchi}},
  \bibinfo {author} {\bibfnamefont {J.}~\bibnamefont {Miyashita}},\ and\
  \bibinfo {author} {\bibfnamefont {M.}~\bibnamefont {{Wada Takeda}}},\
  }\bibfield  {title} {\bibinfo {title} {{Impurity mode in microstrip line
  photonic crystal in millimeter wave region}},\ }\href
  {https://doi.org/10.1143/JPSJ.72.951} {\bibfield  {journal} {\bibinfo
  {journal} {J. Phys. Soc. Jpn.}\ }\textbf {\bibinfo {volume} {72}},\ \bibinfo
  {pages} {951} (\bibinfo {year} {2003})}\BibitemShut {NoStop}%
\bibitem [{\citenamefont {Kitahara}\ \emph {et~al.}(2004)\citenamefont
  {Kitahara}, \citenamefont {Kawaguchi}, \citenamefont {Miyashita},
  \citenamefont {Shimada},\ and\ \citenamefont {{Wada Takeda}}}]{Kitahara2004}%
  \BibitemOpen
  \bibfield  {author} {\bibinfo {author} {\bibfnamefont {H.}~\bibnamefont
  {Kitahara}}, \bibinfo {author} {\bibfnamefont {T.}~\bibnamefont {Kawaguchi}},
  \bibinfo {author} {\bibfnamefont {J.}~\bibnamefont {Miyashita}}, \bibinfo
  {author} {\bibfnamefont {R.}~\bibnamefont {Shimada}},\ and\ \bibinfo {author}
  {\bibfnamefont {M.}~\bibnamefont {{Wada Takeda}}},\ }\bibfield  {title}
  {\bibinfo {title} {{Strongly localized singular Bloch modes created in
  dual-periodic microstrip lines}},\ }\href
  {https://doi.org/10.1143/JPSJ.73.296} {\bibfield  {journal} {\bibinfo
  {journal} {J. Phys. Soc. Jpn.}\ }\textbf {\bibinfo {volume} {73}},\ \bibinfo
  {pages} {296} (\bibinfo {year} {2004})}\BibitemShut {NoStop}%
\bibitem [{\citenamefont {Guo}\ \emph {et~al.}(2008)\citenamefont {Guo},
  \citenamefont {Sun}, \citenamefont {Zhang}, \citenamefont {Li}, \citenamefont
  {Jiang},\ and\ \citenamefont {Chen}}]{Guo2008}%
  \BibitemOpen
  \bibfield  {author} {\bibinfo {author} {\bibfnamefont {J.}~\bibnamefont
  {Guo}}, \bibinfo {author} {\bibfnamefont {Y.}~\bibnamefont {Sun}}, \bibinfo
  {author} {\bibfnamefont {Y.}~\bibnamefont {Zhang}}, \bibinfo {author}
  {\bibfnamefont {H.}~\bibnamefont {Li}}, \bibinfo {author} {\bibfnamefont
  {H.}~\bibnamefont {Jiang}},\ and\ \bibinfo {author} {\bibfnamefont
  {H.}~\bibnamefont {Chen}},\ }\bibfield  {title} {\bibinfo {title}
  {{Experimental investigation of interface states in photonic crystal
  heterostructures}},\ }\href {https://doi.org/10.1103/PhysRevE.78.026607}
  {\bibfield  {journal} {\bibinfo  {journal} {Phys. Rev. E}\ }\textbf {\bibinfo
  {volume} {78}},\ \bibinfo {pages} {026607} (\bibinfo {year}
  {2008})}\BibitemShut {NoStop}%
\bibitem [{\citenamefont {Sasin}\ \emph {et~al.}(2008)\citenamefont {Sasin},
  \citenamefont {Seisyan}, \citenamefont {Kalitteevski}, \citenamefont {Brand},
  \citenamefont {Abram}, \citenamefont {Chamberlain}, \citenamefont {Egorov},
  \citenamefont {Vasil'ev}, \citenamefont {Mikhrin},\ and\ \citenamefont
  {Kavokin}}]{Sasin2008}%
  \BibitemOpen
  \bibfield  {author} {\bibinfo {author} {\bibfnamefont {M.~E.}\ \bibnamefont
  {Sasin}}, \bibinfo {author} {\bibfnamefont {R.~P.}\ \bibnamefont {Seisyan}},
  \bibinfo {author} {\bibfnamefont {M.~A.}\ \bibnamefont {Kalitteevski}},
  \bibinfo {author} {\bibfnamefont {S.}~\bibnamefont {Brand}}, \bibinfo
  {author} {\bibfnamefont {R.~A.}\ \bibnamefont {Abram}}, \bibinfo {author}
  {\bibfnamefont {J.~M.}\ \bibnamefont {Chamberlain}}, \bibinfo {author}
  {\bibfnamefont {A.~Y.}\ \bibnamefont {Egorov}}, \bibinfo {author}
  {\bibfnamefont {A.~P.}\ \bibnamefont {Vasil'ev}}, \bibinfo {author}
  {\bibfnamefont {V.~S.}\ \bibnamefont {Mikhrin}},\ and\ \bibinfo {author}
  {\bibfnamefont {A.~V.}\ \bibnamefont {Kavokin}},\ }\bibfield  {title}
  {\bibinfo {title} {{Tamm plasmon polaritons: Slow and spatially compact
  light}},\ }\href {https://doi.org/10.1063/1.2952486} {\bibfield  {journal}
  {\bibinfo  {journal} {Appl. Phys. Lett.}\ }\textbf {\bibinfo {volume} {92}},\
  \bibinfo {pages} {251112} (\bibinfo {year} {2008})}\BibitemShut {NoStop}%
\bibitem [{\citenamefont {Dyer}\ \emph {et~al.}(2013)\citenamefont {Dyer},
  \citenamefont {Aizin}, \citenamefont {Allen}, \citenamefont {Grine},
  \citenamefont {Bethke}, \citenamefont {Reno},\ and\ \citenamefont
  {Shaner}}]{Dyer2013}%
  \BibitemOpen
  \bibfield  {author} {\bibinfo {author} {\bibfnamefont {G.~C.}\ \bibnamefont
  {Dyer}}, \bibinfo {author} {\bibfnamefont {G.~R.}\ \bibnamefont {Aizin}},
  \bibinfo {author} {\bibfnamefont {S.~J.}\ \bibnamefont {Allen}}, \bibinfo
  {author} {\bibfnamefont {A.~D.}\ \bibnamefont {Grine}}, \bibinfo {author}
  {\bibfnamefont {D.}~\bibnamefont {Bethke}}, \bibinfo {author} {\bibfnamefont
  {J.~L.}\ \bibnamefont {Reno}},\ and\ \bibinfo {author} {\bibfnamefont
  {E.~A.}\ \bibnamefont {Shaner}},\ }\bibfield  {title} {\bibinfo {title}
  {{Induced transparency by coupling of Tamm and defect states in tunable
  terahertz plasmonic crystals}},\ }\href
  {https://doi.org/10.1038/nphoton.2013.252} {\bibfield  {journal} {\bibinfo
  {journal} {Nat. Photonics}\ }\textbf {\bibinfo {volume} {7}},\ \bibinfo
  {pages} {925} (\bibinfo {year} {2013})}\BibitemShut {NoStop}%
\bibitem [{\citenamefont {Xiao}\ \emph {et~al.}(2015)\citenamefont {Xiao},
  \citenamefont {Ma}, \citenamefont {Yang}, \citenamefont {Sheng},
  \citenamefont {Zhang},\ and\ \citenamefont {Chan}}]{Xiao2015}%
  \BibitemOpen
  \bibfield  {author} {\bibinfo {author} {\bibfnamefont {M.}~\bibnamefont
  {Xiao}}, \bibinfo {author} {\bibfnamefont {G.}~\bibnamefont {Ma}}, \bibinfo
  {author} {\bibfnamefont {Z.}~\bibnamefont {Yang}}, \bibinfo {author}
  {\bibfnamefont {P.}~\bibnamefont {Sheng}}, \bibinfo {author} {\bibfnamefont
  {Z.~Q.}\ \bibnamefont {Zhang}},\ and\ \bibinfo {author} {\bibfnamefont
  {C.~T.}\ \bibnamefont {Chan}},\ }\bibfield  {title} {\bibinfo {title}
  {{Geometric phase and band inversion in periodic acoustic systems}},\ }\href
  {https://doi.org/10.1038/nphys3228} {\bibfield  {journal} {\bibinfo
  {journal} {Nat. Phys.}\ }\textbf {\bibinfo {volume} {11}},\ \bibinfo {pages}
  {240} (\bibinfo {year} {2015})}\BibitemShut {NoStop}%
\bibitem [{\citenamefont {Kane}\ and\ \citenamefont {Mele}(2005)}]{Kane2005}%
  \BibitemOpen
  \bibfield  {author} {\bibinfo {author} {\bibfnamefont {C.~L.}\ \bibnamefont
  {Kane}}\ and\ \bibinfo {author} {\bibfnamefont {E.~J.}\ \bibnamefont
  {Mele}},\ }\bibfield  {title} {\bibinfo {title} {{$Z_2$ Topological Order and
  the Quantum Spin Hall Effect}},\ }\href
  {https://doi.org/10.1103/PhysRevLett.95.146802} {\bibfield  {journal}
  {\bibinfo  {journal} {Phys. Rev. Lett.}\ }\textbf {\bibinfo {volume} {95}},\
  \bibinfo {pages} {146802} (\bibinfo {year} {2005})}\BibitemShut {NoStop}%
\bibitem [{\citenamefont {Fu}\ \emph {et~al.}(2007)\citenamefont {Fu},
  \citenamefont {Kane},\ and\ \citenamefont {Mele}}]{Fu2007}%
  \BibitemOpen
  \bibfield  {author} {\bibinfo {author} {\bibfnamefont {L.}~\bibnamefont
  {Fu}}, \bibinfo {author} {\bibfnamefont {C.~L.}\ \bibnamefont {Kane}},\ and\
  \bibinfo {author} {\bibfnamefont {E.~J.}\ \bibnamefont {Mele}},\ }\bibfield
  {title} {\bibinfo {title} {{Topological Insulators in Three Dimensions}},\
  }\href {https://doi.org/10.1103/PhysRevLett.98.106803} {\bibfield  {journal}
  {\bibinfo  {journal} {Phys. Rev. Lett.}\ }\textbf {\bibinfo {volume} {98}},\
  \bibinfo {pages} {106803} (\bibinfo {year} {2007})}\BibitemShut {NoStop}%
\bibitem [{\citenamefont {K{\"{o}}nig}\ \emph {et~al.}(2007)\citenamefont
  {K{\"{o}}nig}, \citenamefont {Wiedmann}, \citenamefont {Br{\"{u}}ne},
  \citenamefont {Roth}, \citenamefont {Buhmann}, \citenamefont {Molenkamp},
  \citenamefont {Qi},\ and\ \citenamefont {Zhang}}]{Konig2007}%
  \BibitemOpen
  \bibfield  {author} {\bibinfo {author} {\bibfnamefont {M.}~\bibnamefont
  {K{\"{o}}nig}}, \bibinfo {author} {\bibfnamefont {S.}~\bibnamefont
  {Wiedmann}}, \bibinfo {author} {\bibfnamefont {C.}~\bibnamefont
  {Br{\"{u}}ne}}, \bibinfo {author} {\bibfnamefont {A.}~\bibnamefont {Roth}},
  \bibinfo {author} {\bibfnamefont {H.}~\bibnamefont {Buhmann}}, \bibinfo
  {author} {\bibfnamefont {L.~W.}\ \bibnamefont {Molenkamp}}, \bibinfo {author}
  {\bibfnamefont {X.-L.}\ \bibnamefont {Qi}},\ and\ \bibinfo {author}
  {\bibfnamefont {S.-C.}\ \bibnamefont {Zhang}},\ }\bibfield  {title} {\bibinfo
  {title} {{Quantum spin Hall insulator state in HgTe quantum wells}},\ }\href
  {https://doi.org/10.1126/science.1148047} {\bibfield  {journal} {\bibinfo
  {journal} {Science}\ }\textbf {\bibinfo {volume} {318}},\ \bibinfo {pages}
  {766} (\bibinfo {year} {2007})}\BibitemShut {NoStop}%
\bibitem [{\citenamefont {Chen}\ \emph {et~al.}(2009)\citenamefont {Chen},
  \citenamefont {Analytis}, \citenamefont {Chu}, \citenamefont {Liu},
  \citenamefont {Mo}, \citenamefont {Qi}, \citenamefont {Zhang}, \citenamefont
  {Lu}, \citenamefont {Dai}, \citenamefont {Fang}, \citenamefont {Zhang},
  \citenamefont {Fisher}, \citenamefont {Hussain},\ and\ \citenamefont
  {Shen}}]{Chen2009}%
  \BibitemOpen
  \bibfield  {author} {\bibinfo {author} {\bibfnamefont {Y.~L.}\ \bibnamefont
  {Chen}}, \bibinfo {author} {\bibfnamefont {J.~G.}\ \bibnamefont {Analytis}},
  \bibinfo {author} {\bibfnamefont {J.-H.}\ \bibnamefont {Chu}}, \bibinfo
  {author} {\bibfnamefont {Z.~K.}\ \bibnamefont {Liu}}, \bibinfo {author}
  {\bibfnamefont {S.-K.}\ \bibnamefont {Mo}}, \bibinfo {author} {\bibfnamefont
  {X.~L.}\ \bibnamefont {Qi}}, \bibinfo {author} {\bibfnamefont {H.~J.}\
  \bibnamefont {Zhang}}, \bibinfo {author} {\bibfnamefont {D.~H.}\ \bibnamefont
  {Lu}}, \bibinfo {author} {\bibfnamefont {X.}~\bibnamefont {Dai}}, \bibinfo
  {author} {\bibfnamefont {Z.}~\bibnamefont {Fang}}, \bibinfo {author}
  {\bibfnamefont {S.~C.}\ \bibnamefont {Zhang}}, \bibinfo {author}
  {\bibfnamefont {I.~R.}\ \bibnamefont {Fisher}}, \bibinfo {author}
  {\bibfnamefont {Z.}~\bibnamefont {Hussain}},\ and\ \bibinfo {author}
  {\bibfnamefont {Z.-X.}\ \bibnamefont {Shen}},\ }\bibfield  {title} {\bibinfo
  {title} {{Experimental realization of a three-dimensional topological
  insulator, Bi$_2$Te$_3$}},\ }\href {https://doi.org/10.1126/science.1173034}
  {\bibfield  {journal} {\bibinfo  {journal} {Science}\ }\textbf {\bibinfo
  {volume} {325}},\ \bibinfo {pages} {178} (\bibinfo {year}
  {2009})}\BibitemShut {NoStop}%
\bibitem [{\citenamefont {Schnyder}\ \emph {et~al.}(2008)\citenamefont
  {Schnyder}, \citenamefont {Ryu}, \citenamefont {Furusaki},\ and\
  \citenamefont {Ludwig}}]{Schnyder2008}%
  \BibitemOpen
  \bibfield  {author} {\bibinfo {author} {\bibfnamefont {A.~P.}\ \bibnamefont
  {Schnyder}}, \bibinfo {author} {\bibfnamefont {S.}~\bibnamefont {Ryu}},
  \bibinfo {author} {\bibfnamefont {A.}~\bibnamefont {Furusaki}},\ and\
  \bibinfo {author} {\bibfnamefont {A.~W.~W.}\ \bibnamefont {Ludwig}},\
  }\bibfield  {title} {\bibinfo {title} {{Classification of topological
  insulators and superconductors in three spatial dimensions}},\ }\href
  {https://doi.org/10.1103/PhysRevB.78.195125} {\bibfield  {journal} {\bibinfo
  {journal} {Phys. Rev. B}\ }\textbf {\bibinfo {volume} {78}},\ \bibinfo
  {pages} {195125} (\bibinfo {year} {2008})}\BibitemShut {NoStop}%
\bibitem [{\citenamefont {Su}\ \emph {et~al.}(1979)\citenamefont {Su},
  \citenamefont {Schrieffer},\ and\ \citenamefont {Heeger}}]{Su1979}%
  \BibitemOpen
  \bibfield  {author} {\bibinfo {author} {\bibfnamefont {W.~P.}\ \bibnamefont
  {Su}}, \bibinfo {author} {\bibfnamefont {J.~R.}\ \bibnamefont {Schrieffer}},\
  and\ \bibinfo {author} {\bibfnamefont {A.~J.}\ \bibnamefont {Heeger}},\
  }\bibfield  {title} {\bibinfo {title} {{Solitons in Polyacetylene}},\ }\href
  {https://doi.org/10.1103/PhysRevLett.42.1698} {\bibfield  {journal} {\bibinfo
   {journal} {Phys. Rev. Lett.}\ }\textbf {\bibinfo {volume} {42}},\ \bibinfo
  {pages} {1698} (\bibinfo {year} {1979})}\BibitemShut {NoStop}%
\bibitem [{\citenamefont {Su}\ \emph {et~al.}(1980)\citenamefont {Su},
  \citenamefont {Schrieffer},\ and\ \citenamefont {Heeger}}]{Su1980}%
  \BibitemOpen
  \bibfield  {author} {\bibinfo {author} {\bibfnamefont {W.~P.}\ \bibnamefont
  {Su}}, \bibinfo {author} {\bibfnamefont {J.~R.}\ \bibnamefont {Schrieffer}},\
  and\ \bibinfo {author} {\bibfnamefont {A.~J.}\ \bibnamefont {Heeger}},\
  }\bibfield  {title} {\bibinfo {title} {{Soliton excitations in
  polyacetylene}},\ }\href {https://doi.org/10.1103/PhysRevB.22.2099}
  {\bibfield  {journal} {\bibinfo  {journal} {Phys. Rev. B}\ }\textbf {\bibinfo
  {volume} {22}},\ \bibinfo {pages} {2099} (\bibinfo {year}
  {1980})}\BibitemShut {NoStop}%
\bibitem [{\citenamefont {Xiao}\ \emph {et~al.}(2014)\citenamefont {Xiao},
  \citenamefont {Zhang},\ and\ \citenamefont {Chan}}]{Xiao2014}%
  \BibitemOpen
  \bibfield  {author} {\bibinfo {author} {\bibfnamefont {M.}~\bibnamefont
  {Xiao}}, \bibinfo {author} {\bibfnamefont {Z.~Q.}\ \bibnamefont {Zhang}},\
  and\ \bibinfo {author} {\bibfnamefont {C.~T.}\ \bibnamefont {Chan}},\
  }\bibfield  {title} {\bibinfo {title} {{Surface Impedance and Bulk Band
  Geometric Phases in One-Dimensional Systems}},\ }\href
  {https://doi.org/10.1103/PhysRevX.4.021017} {\bibfield  {journal} {\bibinfo
  {journal} {Phys. Rev. X}\ }\textbf {\bibinfo {volume} {4}},\ \bibinfo {pages}
  {021017} (\bibinfo {year} {2014})}\BibitemShut {NoStop}%
\bibitem [{\citenamefont {Ashcroft}\ and\ \citenamefont
  {Mermin}(1976)}]{ashcroft1976solid}%
  \BibitemOpen
  \bibfield  {author} {\bibinfo {author} {\bibfnamefont {N.~W.}\ \bibnamefont
  {Ashcroft}}\ and\ \bibinfo {author} {\bibfnamefont {N.~D.}\ \bibnamefont
  {Mermin}},\ }\href {https://books.google.co.jp/books?id=oXIfAQAAMAAJ} {\emph
  {\bibinfo {title} {{Solid State Physics}}}}\ (\bibinfo  {publisher} {Holt,
  Rinehart and Winston},\ \bibinfo {address} {New York},\ \bibinfo {year}
  {1976})\BibitemShut {NoStop}%
\bibitem [{\citenamefont {Vanderbilt}(2018)}]{Vanderbilt2018}%
  \BibitemOpen
  \bibfield  {author} {\bibinfo {author} {\bibfnamefont {D.}~\bibnamefont
  {Vanderbilt}},\ }\href@noop {} {\emph {\bibinfo {title} {{Berry Phases in
  Electronic Structure Theory: Electric Polarization, Orbital Magnetization and
  Topological Insulators}}}}\ (\bibinfo  {publisher} {Cambridge University
  Press},\ \bibinfo {address} {Cambridge, England},\ \bibinfo {year}
  {2018})\BibitemShut {NoStop}%
\bibitem [{\citenamefont {Zak}(1989)}]{Zak1989}%
  \BibitemOpen
  \bibfield  {author} {\bibinfo {author} {\bibfnamefont {J.}~\bibnamefont
  {Zak}},\ }\bibfield  {title} {\bibinfo {title} {{Berry's Phase for Energy
  Bands in Solids}},\ }\href {https://doi.org/10.1103/PhysRevLett.62.2747}
  {\bibfield  {journal} {\bibinfo  {journal} {Phys. Rev. Lett.}\ }\textbf
  {\bibinfo {volume} {62}},\ \bibinfo {pages} {2747} (\bibinfo {year}
  {1989})}\BibitemShut {NoStop}%
\bibitem [{\citenamefont {Vanderbilt}\ and\ \citenamefont
  {King-Smith}(1993)}]{Vanderbilt1993}%
  \BibitemOpen
  \bibfield  {author} {\bibinfo {author} {\bibfnamefont {D.}~\bibnamefont
  {Vanderbilt}}\ and\ \bibinfo {author} {\bibfnamefont {R.~D.}\ \bibnamefont
  {King-Smith}},\ }\bibfield  {title} {\bibinfo {title} {{Electric polarization
  as a bulk quantity and its relation to surface charge}},\ }\href
  {https://doi.org/10.1103/PhysRevB.48.4442} {\bibfield  {journal} {\bibinfo
  {journal} {Phys. Rev. B}\ }\textbf {\bibinfo {volume} {48}},\ \bibinfo
  {pages} {4442} (\bibinfo {year} {1993})}\BibitemShut {NoStop}%
\bibitem [{\citenamefont {Gangadharaiah}\ \emph {et~al.}(2012)\citenamefont
  {Gangadharaiah}, \citenamefont {Trifunovic},\ and\ \citenamefont
  {Loss}}]{Gangadharaiah2012}%
  \BibitemOpen
  \bibfield  {author} {\bibinfo {author} {\bibfnamefont {S.}~\bibnamefont
  {Gangadharaiah}}, \bibinfo {author} {\bibfnamefont {L.}~\bibnamefont
  {Trifunovic}},\ and\ \bibinfo {author} {\bibfnamefont {D.}~\bibnamefont
  {Loss}},\ }\bibfield  {title} {\bibinfo {title} {{Localized End States in
  Density Modulated Quantum Wires and Rings}},\ }\href
  {https://doi.org/10.1103/PhysRevLett.108.136803} {\bibfield  {journal}
  {\bibinfo  {journal} {Phys. Rev. Lett.}\ }\textbf {\bibinfo {volume} {108}},\
  \bibinfo {pages} {136803} (\bibinfo {year} {2012})}\BibitemShut {NoStop}%
\bibitem [{\citenamefont {Park}\ \emph {et~al.}(2016)\citenamefont {Park},
  \citenamefont {Yang}, \citenamefont {Klinovaja}, \citenamefont {Stano},\ and\
  \citenamefont {Loss}}]{Park2016}%
  \BibitemOpen
  \bibfield  {author} {\bibinfo {author} {\bibfnamefont {J.-H.}\ \bibnamefont
  {Park}}, \bibinfo {author} {\bibfnamefont {G.}~\bibnamefont {Yang}}, \bibinfo
  {author} {\bibfnamefont {J.}~\bibnamefont {Klinovaja}}, \bibinfo {author}
  {\bibfnamefont {P.}~\bibnamefont {Stano}},\ and\ \bibinfo {author}
  {\bibfnamefont {D.}~\bibnamefont {Loss}},\ }\bibfield  {title} {\bibinfo
  {title} {{Fractional boundary charges in quantum dot arrays with density
  modulation}},\ }\href {https://doi.org/10.1103/PhysRevB.94.075416} {\bibfield
   {journal} {\bibinfo  {journal} {Phys. Rev. B}\ }\textbf {\bibinfo {volume}
  {94}},\ \bibinfo {pages} {075416} (\bibinfo {year} {2016})}\BibitemShut
  {NoStop}%
\bibitem [{\citenamefont {Thakurathi}\ \emph {et~al.}(2018)\citenamefont
  {Thakurathi}, \citenamefont {Klinovaja},\ and\ \citenamefont
  {Loss}}]{Thakurathi2018}%
  \BibitemOpen
  \bibfield  {author} {\bibinfo {author} {\bibfnamefont {M.}~\bibnamefont
  {Thakurathi}}, \bibinfo {author} {\bibfnamefont {J.}~\bibnamefont
  {Klinovaja}},\ and\ \bibinfo {author} {\bibfnamefont {D.}~\bibnamefont
  {Loss}},\ }\bibfield  {title} {\bibinfo {title} {{From fractional boundary
  charges to quantized Hall conductance}},\ }\href
  {https://doi.org/10.1103/PhysRevB.98.245404} {\bibfield  {journal} {\bibinfo
  {journal} {Phys. Rev. B}\ }\textbf {\bibinfo {volume} {98}},\ \bibinfo
  {pages} {245404} (\bibinfo {year} {2018})}\BibitemShut {NoStop}%
\bibitem [{\citenamefont {Ozawa}\ \emph {et~al.}(2019)\citenamefont {Ozawa},
  \citenamefont {Price}, \citenamefont {Amo}, \citenamefont {Goldman},
  \citenamefont {Hafezi}, \citenamefont {Lu}, \citenamefont {Rechtsman},
  \citenamefont {Schuster}, \citenamefont {Simon}, \citenamefont {Zilberberg},\
  and\ \citenamefont {Carusotto}}]{Ozawa2019}%
  \BibitemOpen
  \bibfield  {author} {\bibinfo {author} {\bibfnamefont {T.}~\bibnamefont
  {Ozawa}}, \bibinfo {author} {\bibfnamefont {H.~M.}\ \bibnamefont {Price}},
  \bibinfo {author} {\bibfnamefont {A.}~\bibnamefont {Amo}}, \bibinfo {author}
  {\bibfnamefont {N.}~\bibnamefont {Goldman}}, \bibinfo {author} {\bibfnamefont
  {M.}~\bibnamefont {Hafezi}}, \bibinfo {author} {\bibfnamefont
  {L.}~\bibnamefont {Lu}}, \bibinfo {author} {\bibfnamefont {M.~C.}\
  \bibnamefont {Rechtsman}}, \bibinfo {author} {\bibfnamefont {D.}~\bibnamefont
  {Schuster}}, \bibinfo {author} {\bibfnamefont {J.}~\bibnamefont {Simon}},
  \bibinfo {author} {\bibfnamefont {O.}~\bibnamefont {Zilberberg}},\ and\
  \bibinfo {author} {\bibfnamefont {I.}~\bibnamefont {Carusotto}},\ }\bibfield
  {title} {\bibinfo {title} {{Topological photonics}},\ }\href
  {https://doi.org/10.1103/RevModPhys.91.015006} {\bibfield  {journal}
  {\bibinfo  {journal} {Rev. Mod. Phys.}\ }\textbf {\bibinfo {volume} {91}},\
  \bibinfo {pages} {015006} (\bibinfo {year} {2019})}\BibitemShut {NoStop}%
\bibitem [{\citenamefont {Asb{\'{o}}th}\ \emph {et~al.}(2016)\citenamefont
  {Asb{\'{o}}th}, \citenamefont {Oroszl{\'{a}}ny},\ and\ \citenamefont
  {P{\'{a}}lyi}}]{asboth2016short}%
  \BibitemOpen
  \bibfield  {author} {\bibinfo {author} {\bibfnamefont {J.~K.}\ \bibnamefont
  {Asb{\'{o}}th}}, \bibinfo {author} {\bibfnamefont {L.}~\bibnamefont
  {Oroszl{\'{a}}ny}},\ and\ \bibinfo {author} {\bibfnamefont {A.}~\bibnamefont
  {P{\'{a}}lyi}},\ }\href@noop {} {\emph {\bibinfo {title} {{A Short Course on
  Topological Insulators}}}}\ (\bibinfo  {publisher} {Springer},\ \bibinfo
  {address} {Cham, Switzerland},\ \bibinfo {year} {2016})\BibitemShut {NoStop}%
\bibitem [{\citenamefont {Thouless}(1983)}]{Thouless1983}%
  \BibitemOpen
  \bibfield  {author} {\bibinfo {author} {\bibfnamefont {D.~J.}\ \bibnamefont
  {Thouless}},\ }\bibfield  {title} {\bibinfo {title} {{Quantization of
  particle transport}},\ }\href {https://doi.org/10.1103/PhysRevB.27.6083}
  {\bibfield  {journal} {\bibinfo  {journal} {Phys. Rev. B}\ }\textbf {\bibinfo
  {volume} {27}},\ \bibinfo {pages} {6083} (\bibinfo {year}
  {1983})}\BibitemShut {NoStop}%
\bibitem [{\citenamefont {Cottey}(1971)}]{Cottey1971}%
  \BibitemOpen
  \bibfield  {author} {\bibinfo {author} {\bibfnamefont {A.~A.}\ \bibnamefont
  {Cottey}},\ }\bibfield  {title} {\bibinfo {title} {{Floquet's theorem and
  band theory in one dimension}},\ }\href {https://doi.org/10.1119/1.1976612}
  {\bibfield  {journal} {\bibinfo  {journal} {Am. J. Phys.}\ }\textbf {\bibinfo
  {volume} {39}},\ \bibinfo {pages} {1235} (\bibinfo {year}
  {1971})}\BibitemShut {NoStop}%
\bibitem [{sup()}]{supplemental}%
  \BibitemOpen
  \href@noop {} {\ }\bibinfo {note} {See Supplemental Material for detailed
  information of the transfer-matrix method, comparison of theoretical and
  experimental transmission data, characterization of localized states, full
  reflection properties of the semi-infinite systems, and winding direction of
  complex reflection amplitudes.}\BibitemShut {Stop}%
\bibitem [{\citenamefont {Collin}(1996)}]{Collin1996}%
  \BibitemOpen
  \bibfield  {author} {\bibinfo {author} {\bibfnamefont {R.~E.}\ \bibnamefont
  {Collin}},\ }\href@noop {} {\emph {\bibinfo {title} {{Foundations for
  Microwave Engineering}}}},\ \bibinfo {edition} {2nd}\ ed.\ (\bibinfo
  {publisher} {McGraw-Hill},\ \bibinfo {address} {New York},\ \bibinfo {year}
  {1996})\BibitemShut {NoStop}%
\bibitem [{\citenamefont {Saleh}\ and\ \citenamefont
  {Teich}(2007)}]{Saleh2007}%
  \BibitemOpen
  \bibfield  {author} {\bibinfo {author} {\bibfnamefont {B.~E.~A.}\
  \bibnamefont {Saleh}}\ and\ \bibinfo {author} {\bibfnamefont {M.~C.}\
  \bibnamefont {Teich}},\ }\href@noop {} {\emph {\bibinfo {title}
  {{Fundamentals of Photonics}}}},\ \bibinfo {edition} {2nd}\ ed.\ (\bibinfo
  {publisher} {John Wiley \& Sons, Inc.},\ \bibinfo {address} {Hoboken, New
  Jersey},\ \bibinfo {year} {2007})\BibitemShut {NoStop}%
\bibitem [{\citenamefont {Hammerstad}\ and\ \citenamefont
  {Jensen}(1980)}]{Hammerstad1980}%
  \BibitemOpen
  \bibfield  {author} {\bibinfo {author} {\bibfnamefont {E.}~\bibnamefont
  {Hammerstad}}\ and\ \bibinfo {author} {\bibfnamefont {O.}~\bibnamefont
  {Jensen}},\ }\bibfield  {title} {\bibinfo {title} {{Accurate models for
  microstrip computer-aided design}},\ }in\ \href
  {https://doi.org/10.1109/MWSYM.1980.1124303} {\emph {\bibinfo {booktitle}
  {1980 IEEE MTT-S Int. Microwave Symp. Dig.}}}\ (\bibinfo  {publisher}
  {IEEE},\ \bibinfo {address} {Washington},\ \bibinfo {year} {1980})\ pp.\
  \bibinfo {pages} {407--409}\BibitemShut {NoStop}%
\end{thebibliography}
%

\clearpage
%
\onecolumngrid
\begin{center}
\textbf{\large Supplemental Material for ``Topological Boundary Modes from Translational Deformations''}
\end{center}
\twocolumngrid

\setcounter{equation}{0}
\setcounter{figure}{0}
\setcounter{table}{0}
\setcounter{page}{1}
\makeatletter
\renewcommand{\theequation}{S\arabic{equation}}
\renewcommand{\thefigure}{S\arabic{figure}}

\renewcommand{\theHtable}{Supplement.\thetable}
\renewcommand{\theHfigure}{Supplement.\thefigure}
\renewcommand{\theHequation}{Supplement.\theequation}



\subsection{Transfer-matrix method}

\begin{figure}[!b]
 \centering
  \includegraphics{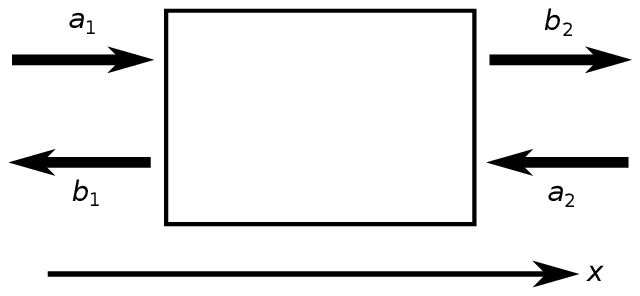}
  \caption{\label{fig:system_model} Model of two-port system.}
 \end{figure}

Here, we explain a calculation technique known as
the transfer-matrix method for 
one-dimensional scalar wave propagation~\cite{Saleh2007,Collin1996}.
Consider a two-port system with
complex amplitudes $a_1$, $a_2$, $b_1$, and $b_2$ of incoming and outgoing signals,
as shown in Fig.~\ref{fig:system_model}.
The scattering property of the system is modeled by a $2\times 2$ scattering matrix $S$ as
\begin{equation}
 \begin{bmatrix}
  b_1 \\
b_2
 \end{bmatrix}
 = S  \begin{bmatrix}
a_1 \\
a_2
 \end{bmatrix}.  \label{eq:8}
\end{equation}
To connect the systems,
we also introduce a $2\times 2$ transfer matrix $T$ as
\begin{equation}
 \begin{bmatrix}
  a_1 \\
b_1
 \end{bmatrix}
 = T  \begin{bmatrix}
b_2 \\
a_2
 \end{bmatrix}.  \label{eq:9}
\end{equation}
By multiplying these transmission matrices,
we obtain a total transmission matrix.
The scattering and transfer matrices are related as follows:
$S_{11} = T_{21}/T_{11}$, $S_{12} = T_{22} - T_{21}T_{12}/T_{11}$,
$S_{21} = 1/T_{11}$, and $S_{22} =-T_{12}/T_{11}$.
Conversely, we have 
$T_{11} = 1/S_{21}$, $T_{12} = -S_{22}/S_{21}$, $T_{21} = S_{11}/S_{21}$, and
$T_{22} = S_{12} - S_{11}S_{22}/S_{21}$.

At a specific point in a one-dimensional photonic system,
we can represent an electric field with an angular frequency $\omega$ as
\begin{equation}
\frac{E}{\sqrt{Z_0 Z_{X}}} = \frac{1}{\sqrt{2}} (a+b) \exp(j \omega t) + \cc,  \label{eq:10}
\end{equation}
where we have an incoming complex amplitude $a$, outgoing complex amplitude $b$,
vacuum impedance $Z_0$, and  relative impedance $Z_{X} = \sqrt{\mu_{X}/\epsilon_{X}}$
with relative permittivity $\epsilon_{{X}}$ and relative permeability $\mu_{{X}}$
at the point in a material $X$. 
Here, $\cc$ represents the complex conjugate of the preceding term.
Note that we use the imaginary unit $j=-i$ for a photonic system,
and $\Im$ represents the coefficient of the imaginary part, which is represented by $j$.
At the boundary between two regions with material $X$ (left) and $Y$ (right),
we have the following scattering matrix:
\begin{equation}
 S^{(b)}_{XY} =  \frac{1}{Z_{X}+Z_{Y}}
\begin{bmatrix}
 Z_{Y}-Z_{X} & 2\sqrt{Z_{X}Z_{Y}}\\
 2\sqrt{Z_{X}Z_{Y}} & Z_{X}-Z_{Y}
\end{bmatrix}.  \label{eq:11}
\end{equation}
Correspondingly, we have $T^{(b)}_{XY}$ as the transfer matrix for $S^{(b)}_{XY}$.
For free propagation across a length $l$ in $X$,
we have the following transfer matrix:
\begin{equation}
 T^{(f)}_{X}(\omega) = 
\begin{bmatrix}
 \exp(j k_{X} l) & 0\\
 0 & \exp(-j k_{X} l)
\end{bmatrix},  \label{eq:12}
\end{equation}
with wave number $k_{X}=\omega/c_{X}$
and speed of light $c_{X}$ in ${X}$.
By using a refractive index $n_{X}=\sqrt{\epsilon_{X}\mu_{X}}$ in ${X}$,
$c_{X}$ is written as $c_0 / n_{X}$, 
with the speed of light $c_0$ in a vacuum. 

Now, we consider a unit cell of a photonic crystal
with length $l_m$ of material $X_m$ from left to right ($m=1,2,...,L$).
The transfer matrix of the unit cell is defined as
\begin{equation}
 T\sur{unit}(\omega) = T^{(b)}_{X_{L} X_{1}} T^{(f)}_{X_1}(\omega) \prod_{m=2}^{L} T^{(b)}_{X_{m-1} X_{m}}
T^{(f)}_{X_m}(\omega).  \label{eq:13}
\end{equation}
Eigenvalues of $T\sur{unit}(\omega)$ are given by $\exp(j k a)$ 
with the complex Bloch wave number $k$ 
and the unit-cell length $a=\sum_m l_m$.
From an eigenvector $\vct{v}=[v_1\ v_2]^\mathrm{T}$ with 
$\Im(k)>0$ (decaying to left) for $T\sur{unit}(\omega)$ inside a band gap,
we can calculate $r_L=  w_1/w_2$ by using
$\vct{w}=[w_1\ w_2]^\mathrm{T}=(T^{(b)}_{X_L V})^{-1}\vct{v}$
with a vacuum $V$ 
for the model shown in Fig.~\ref{fig:reflection_phase_winding}(a) 
of the main text.
The theoretical curves in Figs.~\ref{fig:reflection_phase_winding}(e) and \ref{fig:reflection_phase_winding}(f)
are calculated by this method,
while we regard the wave-launching region as a ``vacuum.''

To obtain the power transmission, we calculate a total transmission matrix $T\sur{tot}$
for the entire system,
and then convert it to the scattering matrix $S\sur{tot}$.
Then, we have the power transmission $|{S\sur{tot}}_{21}|^2$
for an incident wave from the left.
Multiplying a series of transfer matrices to 
$\vct{v}\sub{out} = [{S\sur{tot}}_{21}\ 0]^\mathrm{T}$,
we obtain the electric-field distribution.

\subsection{Comparison of theoretical and experimental transmission data}

\begin{figure}[htbp]
 \centering
  \includegraphics{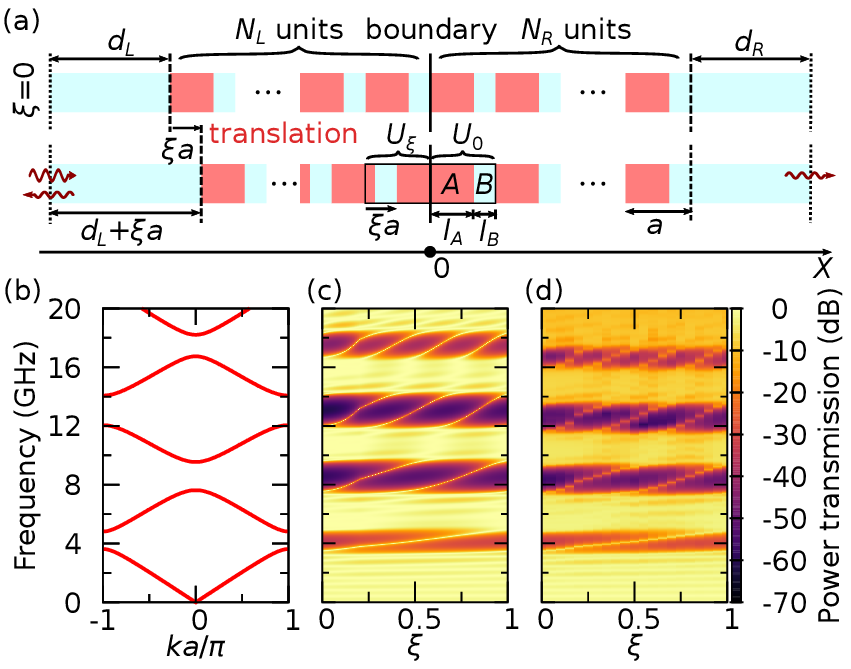}
 \caption{(a)~Theoretical model for the binary photonic crystals shown in 
 Fig.~\ref{fig:exp_trans} of the main text. 
 Starting from a system without a defect (upper), we translate the left region ($x<0$) by $\xi a$ (lower). 
 The structural parameters are $a=12\,\U{mm}$, $l_A/a=0.8$, $l_B/a=0.2$, 
 with $N_L=N_R=5$.
 The regions of $A$ and $B$ have $(n_A, Z_A)=(2.92, 22.6\,\Omega)$ and $(n_B, Z_B)=(2.73, 49.6\,\Omega)$, respectively. Waves enter from the left.
 The two extra regions of $B$ are attached to the system with $d_L/a=d_R/a=2$. 
 (b)~Real band structure of the system without a defect.
 (c)~Theoretically and (d)~experimentally obtained transmission spectra.
 \label{fig:sim_exp_comparison}}
 \vspace{1em}
 \centering
  \includegraphics{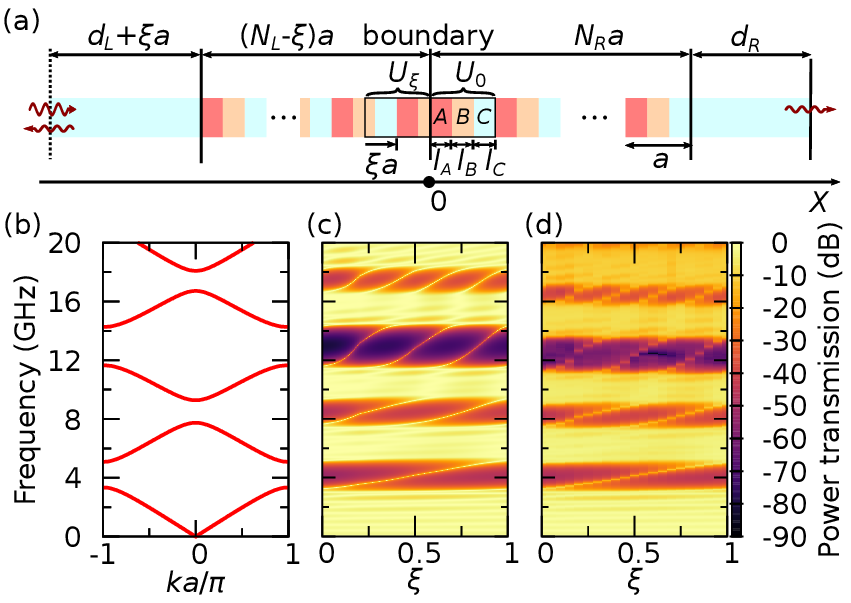}
\caption{(a)~Theoretical model for the ternary photonic crystals shown in 
 Fig.~\ref{fig:exp_trans_without_inversion} of the main text.
The left region ($x<0$) is translated by $\xi a$. 
 The structural parameters are $a=12\,\U{mm}$, $l_A/a=0.5$, $l_B/a=0.3$,
and $l_C/a=0.2$ with $N_L=N_R=5$.
 The regions of $A$, $B$, and $C$ have $(n_A, Z_A)=(2.97, 17.8\,\Omega)$, $(n_B, Z_B)=(2.89, 25.6\,\Omega)$, and $(n_C,\ Z_C)=(2.73, 49.6\,\Omega)$, respectively. Waves enter from the left.
 The two extra regions of $C$ are attached to the system with $d_L/a=d_R/a=2$. 
 (b)~Real band structure of the system without a defect.
 (c)~Theoretically and (d)~experimentally obtained transmission spectra.
 \label{fig:sim_exp_comparison_without_inversion}}
\end{figure}

Here, we calculate transmission spectra of the binary
and ternary photonic crystals [Figs.~\ref{fig:sim_exp_comparison}(a)
and~\ref{fig:sim_exp_comparison_without_inversion}(a), respectively] 
and compare the results with the experimental data. The quasistatic 
refractive indices and impedances of the microstrips 
are calculated from formula derived by E.~Hammerstad and O.~Jensen \cite{Hammerstad1980}.
The calculated photonic bulk bands are shown in Figs.~\ref{fig:sim_exp_comparison}(b) and \ref{fig:sim_exp_comparison_without_inversion}(b), and
the expected transmission spectra are indicated in Figs.~\ref{fig:sim_exp_comparison}(c) and \ref{fig:sim_exp_comparison_without_inversion}(c).
We can see that the bulk bands in Figs.~\ref{fig:sim_exp_comparison}(b)
and~\ref{fig:sim_exp_comparison_without_inversion}(b) correspond to 
the high transmission regions in Figs.~\ref{fig:sim_exp_comparison}(c) 
and \ref{fig:sim_exp_comparison_without_inversion}(c), respectively.
The experimental transmission spectra are also shown in 
Figs.~\ref{fig:sim_exp_comparison}(d) and 
\ref{fig:sim_exp_comparison_without_inversion}(d).
The experimental transmission power is not as high as that calculated in the higher 
frequency region. This is because losses (dielectric, metallic, or radiative) are not included 
in the model. The calculated transmission-band frequencies
show good agreement with the experimental data, especially in the lower 
frequency region. Discrepancies in the higher frequency region can be attributed to the effect of dispersion in the microstrips, 
which is not taken into account in the model calculation. Nonetheless, the 
topological behaviors of the localized states that migrate across 
the transmission gaps
agree very well between theory and experiment in the entire frequency region.

\begin{figure}[!t]
 \centering
  \includegraphics{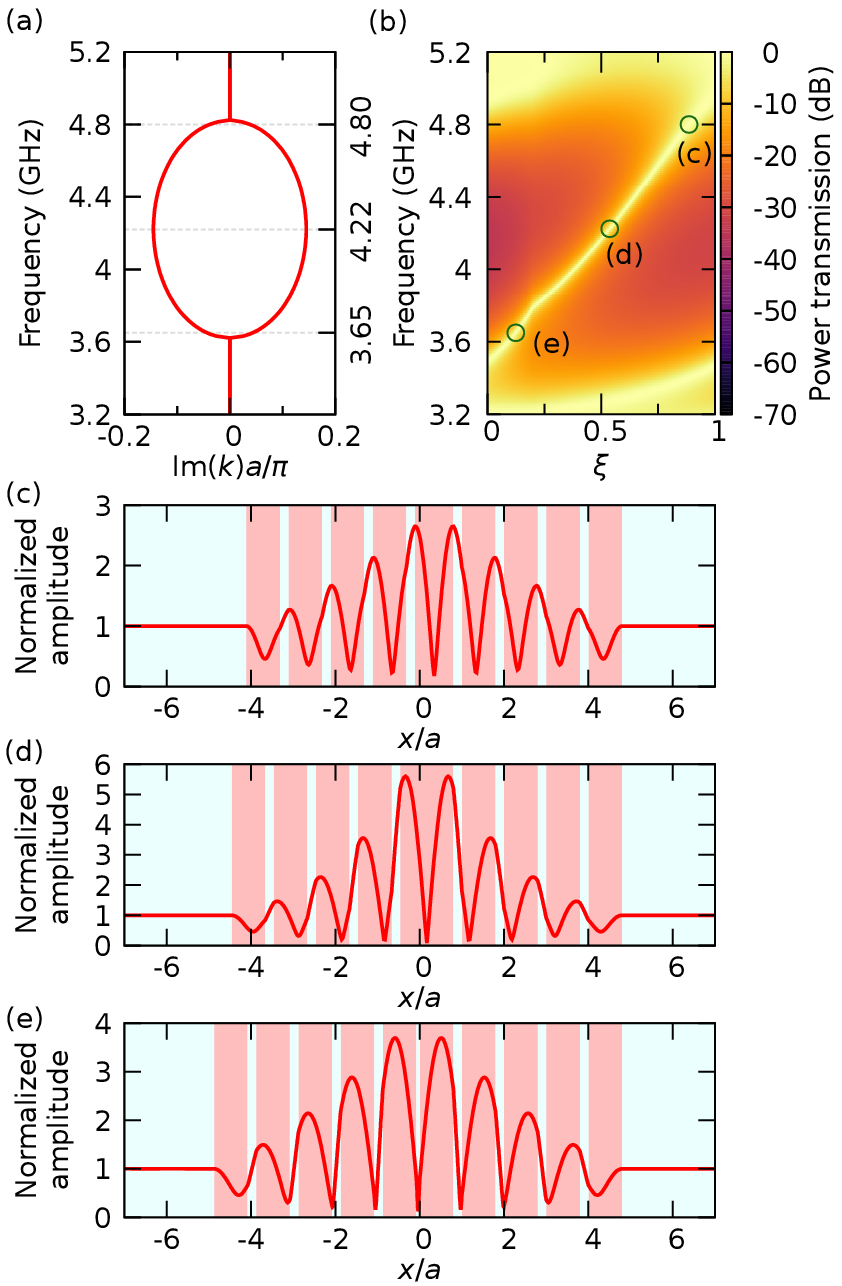}
  \caption{Localization in the first band gap of the binary system 
of Fig.~\ref{fig:sim_exp_comparison} with $N_L=N_R=5$.
(a)~Imaginary part of the complex wave number for the system without a defect.
(b)~Power transmission spectra, indicating an enlarged view of Fig.~\ref{fig:sim_exp_comparison}(c). 
(c)--(e)~Distribution of the absolute value of the complex electric-field amplitude
for  $(\xi, \omega) = (0.887, 2\pi \times 4.8\,\U{GHz})$, 
$(0.538, 2\pi \times 4.22\,\U{GHz})$, and 
$(0.124,2\pi \times 3.65\,\U{GHz})$, respectively. These points are also indicated in (a) by dashed lines and in (b) by circles.
The field values are normalized by the incident-wave amplitude.
 \label{fig:power_trans_and_localization_detail}}
 \end{figure}

\subsection{Characterization of localized states}

\begin{figure}[!t]
 \centering
  \includegraphics{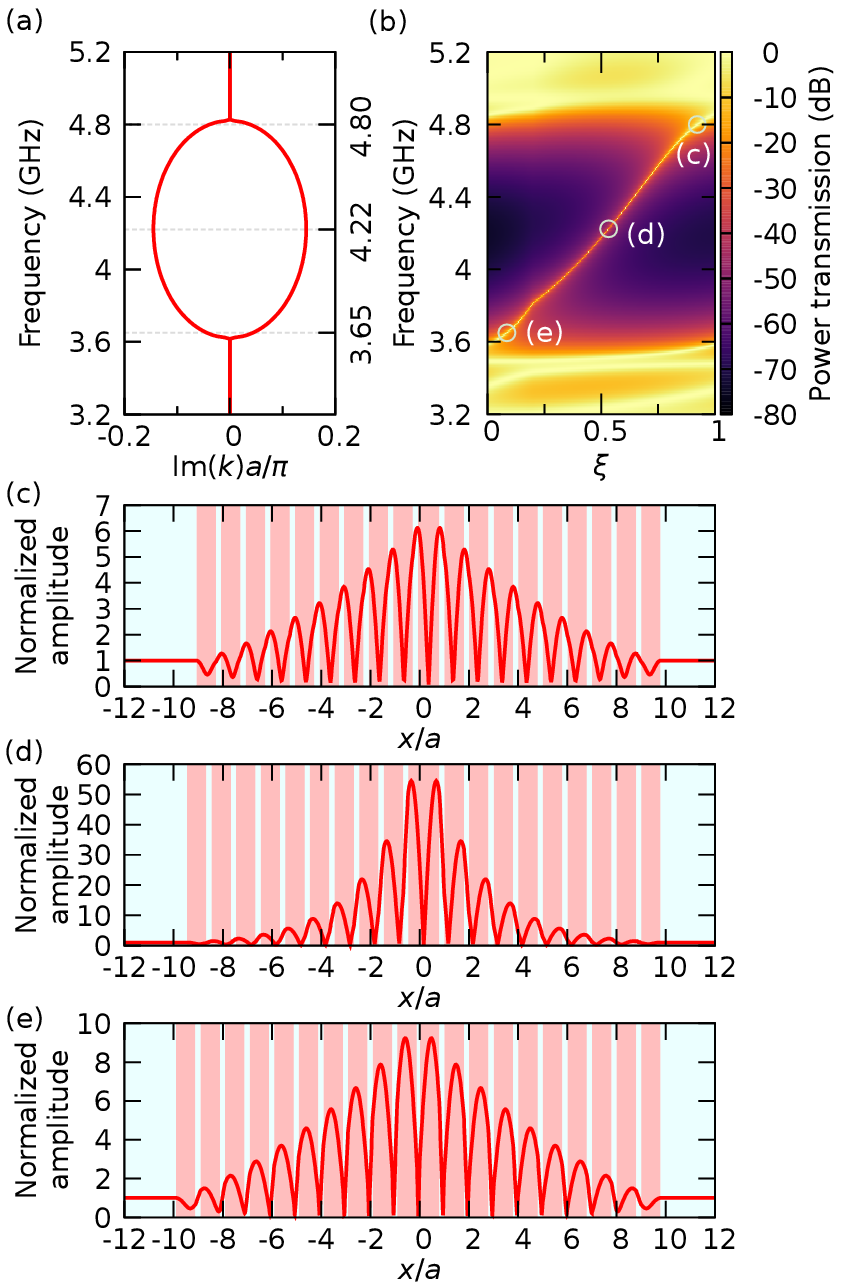}
  \caption{Localization in the first band gap of the binary system 
with $N_L=N_R=10$. The other parameters are the same as those in
Fig.~\ref{fig:sim_exp_comparison}.
(a) Imaginary part of the complex wave number for the system without a defect. 
(b) Calculated power transmission spectra.
(c)--(e) Distribution of the absolute value of the complex electric-field amplitude
for $(\xi, \omega) = (0.923, 2\pi \times 4.80\,\U{GHz})$, $(0.533, 2\pi \times 4.22\,\U{GHz})$, and 
$(0.0855,2\pi \times 3.65\,\U{GHz})$. These points are also indicated in (a) by dashed lines and in (b) by circles.
The field values are normalized by the incident-wave amplitude.
\label{fig:power_trans_and_localization_longer_detail} }
 \end{figure}

\begin{figure}[!t]
 \centering
  \includegraphics{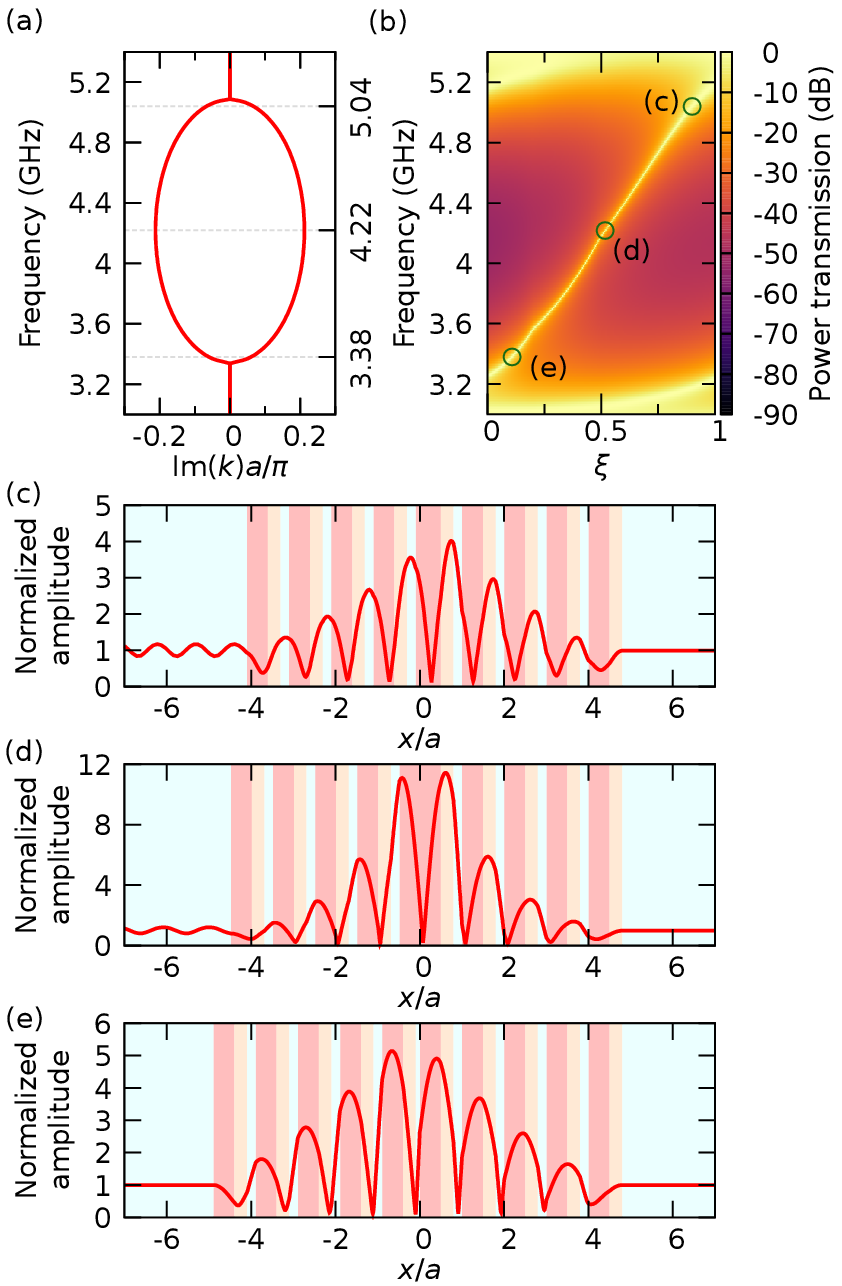}
  \caption{Localization in the first band gap of the ternary system 
of Fig.~\ref{fig:sim_exp_comparison_without_inversion} with $N_L=N_R=5$.
(a)~Imaginary part of the complex wave number for the system without a defect. 
(b)~Power transmission spectra, indicating an enlarged view of Fig.~\ref{fig:sim_exp_comparison_without_inversion}(c). 
(c)--(e) Distribution of the absolute value of the complex electric-field amplitude
for $(\xi, \omega) = (0.899, 2\pi \times 5.04\,\U{GHz})$, $(0.517, 2\pi \times 4.22\,\U{GHz})$, and 
$(0.107,2\pi \times 3.38\,\U{GHz})$, respectively. These points are also indicated in (a) by dashed lines and in (b) by circles.
The field values are normalized by the incident-wave amplitude.
 \label{fig:power_trans_and_localization_without_inversion_detail}}
 \end{figure}

In this section, we characterize the field distribution of the localized states,
based on the theoretical-model calculations.
First, we analyze the binary photonic crystals.
Generally, we have complex wave numbers inside band gaps.
Figure~\ref{fig:power_trans_and_localization_detail}(a) 
shows the imaginary part of the complex wave number inside the first band gap
for the model.
Near the center of the band gap, 
the imaginary part is maximized.
Figure~\ref{fig:power_trans_and_localization_detail}(b) 
shows the power transmission, which is enlarged from 
Fig.~\ref{fig:sim_exp_comparison}(c).
To see the variation of the distribution,
we took three points along the boundary-mode dispersion
depicted as circles in Fig.~\ref{fig:power_trans_and_localization_detail}(b).
The corresponding field distributions are plotted in 
Figs.~\ref{fig:power_trans_and_localization_detail}(c)--(e).
As we expected from Fig.~\ref{fig:power_trans_and_localization_detail}(a), 
localization is the narrowest in the center of the band gap.
To observe the localization tuning more clearly, 
we increase $N_L=N_R=10$ from $N_L=N_R=5$
while the other parameters are left unchanged.
The calculated results are summarized in 
Fig.~\ref{fig:power_trans_and_localization_longer_detail}.
A comparison of Figs.~\ref{fig:power_trans_and_localization_longer_detail}(c)--(e) clearly shows the realization of the narrowest localization in the band-gap center.
Thus, we can tune localization by altering $\xi$, i.e., the termination.
The narrowest localization decreases the coupling to the incident wave,
and the radiative loss is reduced more effectively.
For completeness,
we also show data for the ternary photonic crystals
in Fig.~\ref{fig:power_trans_and_localization_without_inversion_detail}.
The inversion-symmetry breaking leads to nonsymmetric 
field distributions in Figs.~\ref{fig:power_trans_and_localization_without_inversion_detail}(c)--(e).

\subsection{Full reflection properties of the half systems}

\begin{figure}[!t]
 \centering
  \includegraphics{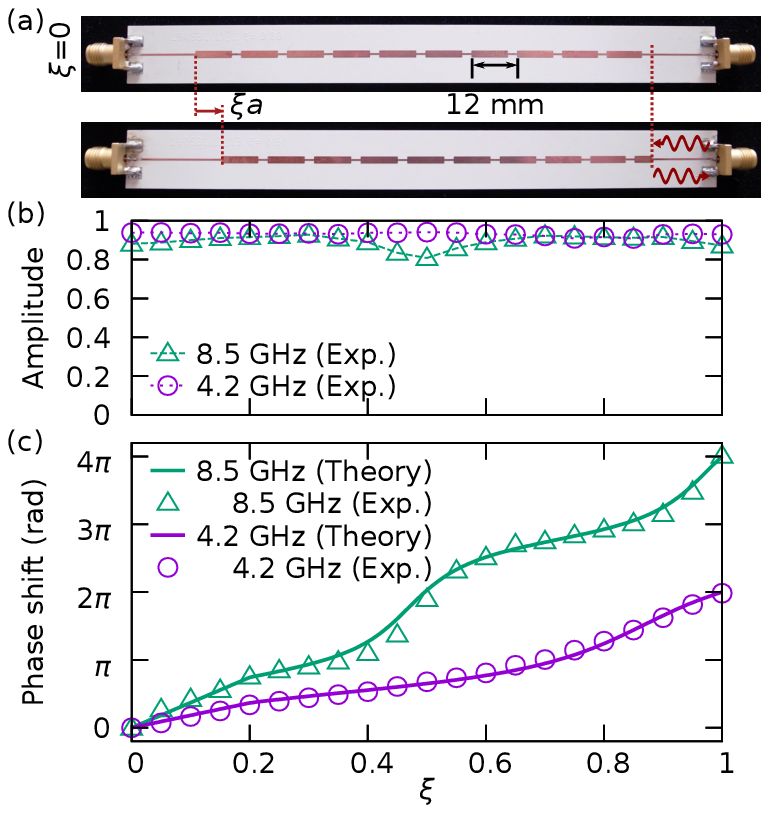}
  \caption{
(a)~Photograph of binary samples ($\xi=0$ and $0.60$) to measure $r_L(\xi,\omega)$.
(b) Reflection amplitude $|r_L(\xi, \omega_0)|$
and (c) phase shift $\arg[r_L(\xi, \omega_0)] - \arg[r_L(0, \omega_0)]$
as a function of $\xi$.
Here, $\omega_0$ is set to
$2\pi\times 4.20\,\U{GHz}$ and
$2\pi\times 8.50\,\U{GHz}$
inside the first and second band gaps, respectively.
A microwave is injected from the right connector; meanwhile,
the left connector is connected to 
another port of the network analyzer through a cable.
The theoretical curves for the semi-infinite system are plotted in (c) with the experimentally obtained points.
 \label{fig:reflection_phase_winding_detail}}
 \end{figure}

\begin{figure}[!t]
 \centering
  \includegraphics{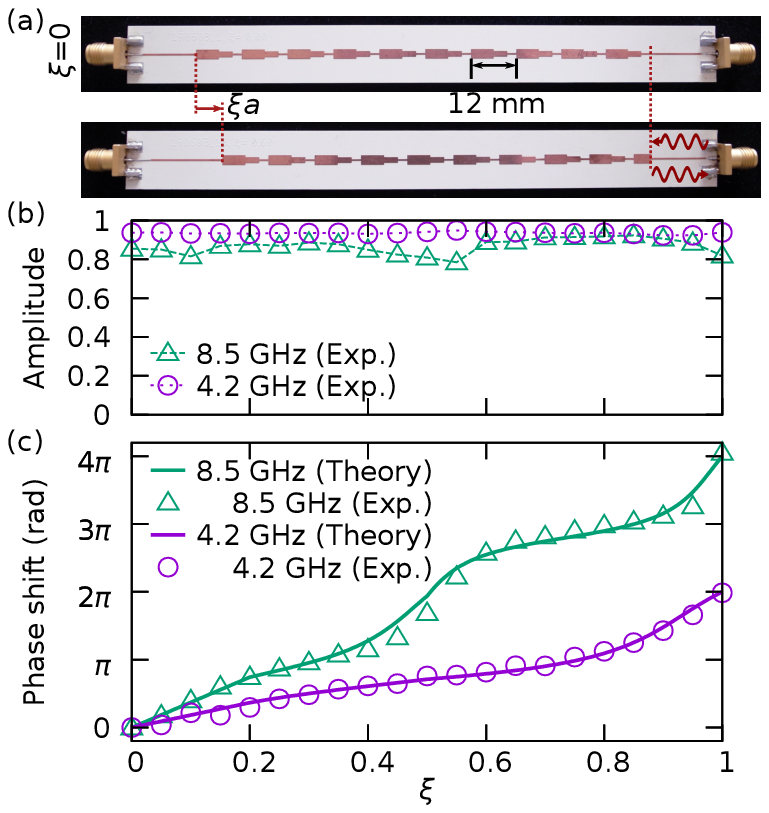}
  \caption{
(a)~Photograph of ternary samples ($\xi=0$ and $0.60$) 
to measure $r_L(\xi,\omega)$.
(b)~Reflection amplitude $|r_L(\xi, \omega_0)|$
and (c)~phase shift $\arg[r_L(\xi, \omega_0)] - \arg[r_L(0, \omega_0)]$
as a function of $\xi$.
Here, $\omega_0$ is set to
$2\pi\times 4.20\,\U{GHz}$ and
$2\pi\times 8.50\,\U{GHz}$
inside the first and second band gaps, respectively.
A microwave is injected from the right connector; meanwhile,
the left connector is connected to 
another port of the network analyzer through a cable.
The theoretical curves for the semi-infinite system 
are plotted in (c) with the experimentally obtained points.
 \label{fig:reflection_phase_winding_without_inversion_detail}}
 \end{figure}

Here, we provide reflection-amplitude data for the 
half photonic crystals,
in addition to the phase data of Figs.~\ref{fig:reflection_phase_winding}(e) and \ref{fig:reflection_phase_winding}(f).
Figures~\ref{fig:reflection_phase_winding_detail} and
\ref{fig:reflection_phase_winding_without_inversion_detail}
show the complete data of reflection coefficients
for the binary and ternary halves, respectively.
The reflection amplitude $|r_L(\xi, \omega_0)|$
should be unity  in the theoretical models; however, 
it is degraded by the finite dissipation in the experiments.
Slight changes in the reflection amplitudes with respect to $\xi$
are observed in Figs.~\ref{fig:reflection_phase_winding_detail}(b) and
\ref{fig:reflection_phase_winding_without_inversion_detail}(b).
This can be attributed to the resonance, 
which is caused by a change in the length of 
the first strip located near the reflection boundary.

\subsection{Winding direction of complex reflection amplitudes}

\begin{figure}[!t]
 \centering
  \includegraphics{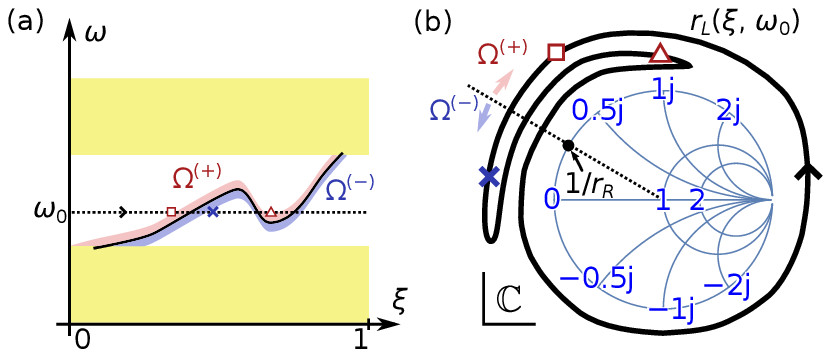}
  \caption{(a)~Possible situation for an boundary-mode transition 
for the first band gap. (b)~Corresponding trajectory of $r_L(\xi, \omega_0)$
from $\xi=0$ to $1$ with the Smith chart, while fixing $\omega = \omega_0$.
 \label{fig:reflection_winding_proof}}
 \end{figure}

Based on Foster's theorem, let us determine the rotation direction of 
$r_L(\xi,\omega_0)$
from $\xi=0$ to $\xi=1$ for fixed $\omega_0$ inside a band gap.
In this section, we focus on the band gap 
between the $n$th and $(n+1)$th bands.
Thus, the total number of migrating localized modes 
from the $n$th to $(n+1)$th bands is $n$.

The eigenfrequencies of localized modes inside the band gap
are written as $\omega=\omega_m(\xi)$, which are 
determined by $r_L(\xi,\omega) r_R(\omega)=1$.
Here, we consider the following regions with a small $\Delta \omega>0$: 
$\Omega^{(+)} = \bigcup_m \Omega^{(+)}_m$
and $\Omega^{(-)} = \bigcup_m \Omega^{(-)}_m$,
where $\Omega^{(+)}_m=\{(\xi, \omega)| \xi\in [0,1],\, 
\omega_m(\xi) < \omega < \omega_m(\xi) + \Delta \omega\}$
and $\Omega^{(-)}_m=\{(\xi, \omega)| \xi\in [0,1],\, 
\omega_m(\xi) - \Delta \omega < \omega < \omega_m(\xi)\}$.
A possible situation for $n=1$ is graphically 
shown in Fig.~\ref{fig:reflection_winding_proof}.
Foster's reactance theorem can be applied,
provided that the system is passive~\cite{Collin1996}.
Then, $r_L(\xi,\omega) r_R(\omega)$
must monotonically rotate 
clockwise in the complex plane when we increase $\omega$.
Therefore, we have $\arg r_L < -\arg r_R$
and $\arg r_L > -\arg r_R$
for $\Omega^{(+)}$ and $\Omega^{(-)}$, respectively.

Now, consider $r_L(\xi,\omega_0)$, 
with a specified $\omega_0$ inside the band gap.
We assume there is no degeneracy of localized states
at $\omega=\omega_0$.
In other words, we always have $p=q$ 
if $\omega_p(\xi) = \omega_q(\xi)=\omega_0$.
By changing $\xi$ from 0 to 1 along $\omega=\omega_0$,
the total number of transitions from $\Omega^{(+)}$ to $\Omega^{(-)}$ 
must be $n$ owing to the bulk-edge correspondence.
In the Smith chart, $\Omega^{(+)} \rightarrow \Omega^{(-)}$
corresponds to the situation that $r_L$ crosses over 
$[r_R(\omega_0)]^{-1}$ in 
an anti-clockwise manner. Similarly, $\Omega^{(-)}\rightarrow \Omega^{(+)}$  
represents clockwise crossing.
Then, the winding number in the Smith chart must be $n$ 
in an anti-clockwise manner with changing 
$\xi$ from $0$ to $1$ because the trajectory is continuous.
If there is an accidental degeneracy, we may consider an angular frequency 
$\omega_0'$ that is slightly displaced from $\omega_0$ to avoid this degeneracy.
Owing to the continuity, 
the winding number at $\omega_0$ must be the same as that at $\omega_0'$.

\end{document}